\documentclass[aps,showpacs,prd,twocolumn,amsmath,amssymb,nofootinbib,letterpaper,floatfix]{revtex4}
\pdfoutput=1
\usepackage{graphicx}
\usepackage{epstopdf}
\graphicspath{{figs/}}
\usepackage{tabularx}
\usepackage{multirow}
\usepackage{wrapfig}
\usepackage{setspace}
\usepackage{tikz}
\usetikzlibrary{arrows,shapes}
\usetikzlibrary{decorations.markings}
\usetikzlibrary{shapes,snakes}
\usetikzlibrary{calc}
\usepackage{footmisc}
\makeatletter
\renewcommand{\@makefntext}[1]{\parindent=1em\noindent\hbox to 1.8em{\hss$^{\@thefnmark}$}#1}
\renewcommand{\@footnotemark}{\hbox{\mathsurround=0pt$^{\@thefnmark}$}}

\newcommand{\E}{\mathrm{e}}
\newcommand{\QCDbar}{QCD$_\textrm{r}$ }
\makeatother

\begin{document}
\title{Isoscalar mesons upon unbreaking of chiral symmetry}
\author{M. Denissenya}
\email{mikhail.denissenya@uni-graz.at}
\author{ L.Ya. Glozman }
\email{leonid.glozman@uni-graz.at}
\author{ C.B. Lang }
\email{christian.lang@uni-graz.at}
\affiliation{Institute of Physics,  University of Graz, A--8010 Graz, Austria}
\begin{abstract}
In a dynamical lattice simulation with the overlap Dirac operator and $N_f=2$ mass degenerate quarks
we study all
possible  $J=0$ and $J=1$ correlators upon
exclusion of the low lying ``quasi-zero'' modes from the valence quark propagators.
After  subtraction of a   small amount of such Dirac
eigenmodes all disconnected contributions vanish and
all possible point-to-point $J=0$ correlators with different quantum
numbers become identical, signaling a restoration of the 
$SU(2)_L \times SU(2)_R \times U(1)_A$.  The original ground state of  the
$\pi$ meson
does not survive this truncation, however.  In contrast, in the
$I=0$  and $I=1$   channels for the $J=1$ correlators the ground states have a very 
clean exponential decay. All possible chiral multiplets for the $J=1$ mesons become degenerate, 
indicating a restoration of an $SU(4)$ symmetry of the dynamical QCD-like string.  
\end{abstract}

\maketitle
\section{ Introduction} 

In Ref. \cite{DGL} we have studied  the behavior of masses of the
isovector $J=1$ mesons $\rho,\rho',a_1,b_1$ upon subtraction of the
lowest-lying eigenmodes of the manifestly chirally-invariant overlap
Dirac operator \cite{OVERLAP} from the valence quark propagators (for
a previous lattice study with a chirally-improved Dirac operator
see Refs. \cite{LS,GLS}). A non-vanishing density of the quasi-zero modes of the Dirac
operator (in the infinite volume limit) represents, through the Banks-Casher relation \cite{BC}, the quark
condensate of the vacuum. Consequently, a removal of a sufficient amount
of the lowest-lying Dirac eigenmodes should eventually lead to the artificial
restoration ("unbreaking") of  chiral symmetry. 

Of course,  correlators
obtained after such a truncation do not correspond to a  local quantum field
theory  \cite{myfootnote}  and we call that case \QCDbar
in order to  distinguish it from the full, untruncated QCD. Despite that fact the correlators shown
intriguing behavior.
Firstly, they have  a very clean exponential decay
for all isovector $J=1$ mesons, suggesting that these states survive the unbreaking
procedure. Secondly, they have interesting symmetry patterns: All
$\rho,\rho',a_1,b_1$  states become degenerate. Certainly this degeneracy is not
accidental and tells us something important about the underlying dynamics.
From this degeneracy
we could infer a simultaneous restoration of both $SU(2)_L \times SU(2)_R$ and $U(1)_A$ symmetries.
However, a degeneracy of all isovector states implies a larger symmetry that includes
$SU(2)_L \times SU(2)_R$ and $U(1)_A$ as subgroups. This larger symmetry would require a
degeneracy of all possible chiral multiplets that contain both
isovector and isoscalar $J=1$ mesons. One of the principal
purposes of the present study is to investigate the isoscalar
$J=1$ states from all possible chiral multiplets 
and clarify whether they 
become degenerate upon unbreaking of the chiral symmetry.

This larger symmetry has been identified in Ref. \cite{G} as an $SU(4) \supset SU(2)_L \times SU(2)_R \times U(1)_A$,
 that mixes 
components of the fundamental four-component vector $ (u_L,u_R,d_L,d_R)$. This symmetry,
which is a symmetry of the confining interaction in \QCDbar,
is not a symmetry of the original QCD Lagrangian and should be considered as an emergent symmetry
that appears from the QCD dynamics upon subtraction of the quasi-zero modes of the Dirac
operator. From the degeneracy of all possible $J=1$ chiral multiplets and from this symmetry 
it was possible to conclude
that there is no color-magnetic field in the system suggesting that we observe quantum
levels of a dynamical \QCDbar string.

Our second aim in the present paper is to study  the fate of the ground states of 
the $\pi, \sigma, a_0, \eta$ mesons upon unbreaking of the chiral symmetry. One
naturally
expects a disappearance of the pion as a Goldstone boson from the spectrum. However, 
a priori it is not clear what behavior to expect for the other $J=0$ mesons. 
We find that the disconnected contributions vanish and that the point-to-point
correlation functions in all $J=0$ channels become indistinguishable
after removal of the lowest-lying modes.
This confirms a restoration of the
$SU(2)_L \times SU(2)_R$ and $U(1)_A$ symmetries. The eigenvalues of the correlation matrices
that  correspond to the ground states of $\pi, \sigma, a_0, \eta$ mesons loose, however, the
exponential decay property implying that the unbreaking removes the physical ground states of
 $\pi, \sigma, a_0, \eta$ mesons from the spectrum. 

\section{Lattice techniques}
\subsection{Quark propagators}

As was discussed in the Introduction we use in our study the overlap Dirac operator. 
The gauge ensemble consists of 100 gauge  configurations generated with $N_f=2$
mass degenerate dynamical overlap fermions on a $16^3\times 32$ lattice with a lattice spacing $a \sim 0.12$ fm
\cite{KEK,Aoki:2012pma}. The  gauge configurations were generated by
constraining the simulation always to the same topological sector $Q_T=0$. The
effect of this was discussed in Ref. \cite{Aoki:2007ka}. While it is known that there
is a slight dependence $O(1/V)$ of the bound state masses on the
topological sector \cite{Brower:2003yx} it is expected to be very small \cite{Bali:2001gk} compared to
our statistical error and should become irrelevant in the infinite volume 
limit.\footnote{In our previous study without fixing topological charge \cite{GLS} the
observed symmetries were noted as well, although with less precision.}

The pion mass in this ensemble is
$M_{\pi} =289(2)$ MeV \cite{Noaki:2008iy}. The overlap quark propagators
and gauge configurations were generously provided by the JLQCD collaboration 
\cite{KEK,Aoki:2012pma,Noaki:2008iy}. 
The quark propagators were computed by the combining
of the exact 100 low modes (eigenvectors of the Dirac operator) with the stochastic estimates for the higher modes, 
(see Appendix A for details). 

To unbreak the chiral symmetry we exclude the lowest-lying modes from the valence quark propagators. 
We introduce the reduced valence quark propagators using the spectral representation where an
increasing number $k$ of the near-zero modes  is excluded from the full propagator:
 \begin{equation}\label{def_trunc}
  S_{k}(x,y)=\sum^{100}_{n=k+1}\frac{1}{\lambda_n}u_n(x)u_n^\dagger(y)+S_{Stoch}
 \end{equation} 
 Full (not reduced) quark propagators correspond to  $k=0$ and all reduced to $k>0$. 

\subsection{Meson observables}

Meson spectroscopy is done by means of the standard variational approach
\cite{VAR}. In each channel we use a set of interpolators  with all possible chiral structures
\cite{COHEN}, see Table \ref{tab:int}.
Each of these operators is constructed from quark propagators with smeared sources.
The exponential type of smearing with a set of different smearing parameters is used
at the source/sink and summarized in Appendix A (for the
computational techniques see \cite{Aoki:2009qn,Aoki:2012pma}). This way we have several
interpolators in each quantum channel and can use the variational method.

After construction of  the cross-correlation matrices  
\begin{equation}
C_{ij}(t)=\langle 0|\mathcal{O}_i(t)\mathcal{O}_j^\dagger(0)|0\rangle
\end{equation}
with the size of the matrices $C_{ij}$ ranging from $5\times5$ to $10\times10$ we solve the
generalized eigenvalue problem  
\begin{equation}
C(t)\vec{\upsilon}_n(t)=\tilde{\lambda}^{(n)}(t)C(t_0)\vec{\upsilon}_n(t)\;.
\end{equation}
Resulting masses of the states are obtained by identifying the exponential behavior of
the eigenvalues $\tilde{\lambda}^{(n)}(t,t_0)= \E^{-E_n (t-t_0)}\left(1+
\mathcal{O}\left(\E^{-\Delta E_n (t-t_0)}\right) \right)$ with $t_0=1$.  
Such states are extracted for each  truncation level $k$ in a given quantum channel.

\begin{table}[h]
\begin{center}
\begin{ruledtabular}
\begin{tabular}{| c | c | c |} 
$I,J^{PC}$ & $\mathcal{O}$ & $R$  \\ \hline
$\pi\;(1,0^{-+})$ & $\bar{q}\gamma_5 \frac{\vec\tau}{2}q$& $(1/2,1/2)_a$\\
$\eta\;(0,0^{-+})$ &  $\bar{q}\gamma_5 q$ & $(1/2,1/2)_b$\\   
$a_0(1,0^{++})$ &  $\bar{q}\frac{\vec\tau}{2} q$ & $(1/2,1/2)_b$ \\
$\sigma\;(0,0^{++})$ &  $\bar{q} q$   & $(1/2,1/2)_a$            \\
 \hline
\multicolumn{1}{|c|}{\multirow{2}{20mm}{\hfil$\rho(1,1^{--})$}} 
&$\bar{q}\gamma_i\frac{\vec\tau}{2}q$& $(1, 0) + (0, 1)$\\ 
                         &      $\bar{q}\gamma_i\gamma_t\frac{\vec\tau}{2}q$& $(1/2,1/2)_b$  \\
\hline
\multicolumn{1}{|c|}{\multirow{2}{20mm}{\hfil$\omega(0,1^{--})$}}&  $\bar{q}\gamma_iq$&
$(0,0)$ \\ 
                         &      $\bar{q}\gamma_i\gamma_t q$ & $(1/2,1/2)_a$ \\ \hline
        $a_1(1,1^{++})$  &   $\bar{q}\gamma_i \gamma_5\frac{\vec\tau}{2}q$ & $(1, 0) + (0,
1)$  \\ 
        $f_1(0,1^{++})$  & $\bar{q}\gamma_i\gamma_5q $   & $(0,0)$  \\ 
        $b_1(1,1^{+-})$  &   $\bar{q} \gamma_i \gamma_j \frac{\vec\tau}{2}q$ & $(1/2,1/2)_a$ \\ 
        $h_1(0,1^{+-})$  &  $\bar{q} \gamma_i\gamma_jq$ & $(1/2,1/2)_b$                  \\  
 \end{tabular}
\end{ruledtabular}
 \end{center}
 \caption{$J=0$ and $J=1$ meson interpolating fields with the corresponding chiral
representation $R$.}\label{tab:int}
\end{table}

\section{Results}
\subsection{ $J=0$ mesons}

The symmetry connections between the $J=0$ interpolators \cite{Glozman:2007ek}
are shown  in Fig. \ref{fig:symj0}.
If both the $SU(2)_L \times SU(2)_R$ and $U(1)_A$ symmetries get restored
we should observe, on the one hand, a coincidence of the point-to-point correlators
obtained with all these four operators. On the other hand, if there are physical states
with the corresponding quantum numbers, i.e., $\pi, \sigma, a_0, \eta$, they must be degenerate.
One should stress that a coincidence of the point-to-point correlators does not imply yet existence
of the degenerate physical states. Only if the eigenvalues of the correlation matrices show an
exponential
decay at large time one can speak about the states.

\begin{figure}[h]
\begin{center}
\begin{tikzpicture}[<->,>=stealth',shorten >=1pt,auto,
                    thick,font=\small,main
node/.style={rectangle,draw=none,text width=6cm,text centered,font=\sffamily\bfseries}]
  \node[node distance=4cm] (1) { $\pi$($\bar{q}\gamma_5 \frac{\vec\tau}{2}q$)};
  \node[node distance=4cm] (2) [right of=1] { $\sigma$($\bar{q}q$)$\quad\,$ };
  \node[node distance=1.1cm] (3) [below of=2] { $\eta$($\bar{q}\gamma_5 q$)$\,\,$};
  \node[node distance=1.1cm] (4) [below of=1] { $a_0$($\bar{q} \frac{\vec\tau}{2}q$)$\,\,\,\,\,$};
   \node[node distance=2cm] (5) [left of=1] {$(1/2,1/2)_a$: };
     \node[node distance=2cm] (6) [left of=4] {$(1/2,1/2)_b$: }; 
  \path[every node/.style={font=\sffamily}]
    (1) edge [] node[ pos=0.5, sloped, above] {\small $SU(2)_L\times SU(2)_R$}(2)
    (2) edge [] node[ pos=0.5,  right] {\small $U(1)_A$}                       (3)
    (3) edge [] node[ pos=0.5, sloped, below] {\small $SU(2)_L\times SU(2)_R$}(4)
    (4) edge [] node[ pos=0.5, left] {\small $U(1)_A$}                         (1);
\end{tikzpicture}
 \end{center}
\caption{Symmetry relations among $J=0$ interpolators.}\label{fig:symj0}
\end{figure}
   
\begin{figure*}[t!]
\tikzset{middlearrow/.style={
        decoration={markings,
            mark= at position 0.5 with {\arrow{#1}} ,
        },
        postaction={decorate}
    }
}
 \begin{tikzpicture}
\draw [middlearrow={latex }][ultra thick,black] (-1.9,-0.86) to[out=60,in=185] (-1,0)
to[out=-15,in=155] (0.5,0) to[out=-15,in=160] (1.86,-0.9);
\node (1) [draw=black,thick,circle,minimum size=0.1cm]  at (-2,-1)  {};
\node (2) [draw=black,fill=black,thick,circle, minimum size=0.1cm]  at (2,-1)  {};
\draw [middlearrow={latex reversed}][ultra thick,black] (-2,-1.5) to[out=-60,in=-185] (-1,-2)
to[out=15,in=-155] (0.75,-2)
to[out=15,in=-160] (1.86,-1.6);
\node (3) [draw=black,fill=black,thick,circle,minimum size=0.1cm]  at (-2,-1.5)  {};
\node (4) [draw=black,thick,circle, minimum size=0.1cm]  at (2,-1.5)  {};
\node (5) [draw=black,thick,rotate=90,dashed,ellipse,minimum height=20pt,minimum width=35pt] 
at (-2,-1.25) {};
\node (6) [draw=black,thick,rotate=90,dashed,ellipse,minimum height=20pt,minimum width=35pt] at
(2,-1.25) {};
\node at (-2.6,-1.25) {0};
\node at (2.6,-1.25) {$x$};
\node (5) [draw=black,thick,rotate=90,dashed,ellipse,minimum height=20pt,minimum width=35pt] 
at (4,-1.25) {};
\draw [middlearrow={latex }][ultra thick,black] (4.12,-0.9) to[out=30,in=185] (5,0)
to[out=0,in=0] (5,-2.5) to[out=175,in=0] (4,-1.5);
\node  [draw=black,thick,circle,minimum size=0.1cm]  at (4,-1)  {};
\node  [draw=black,fill=black,thick,circle,minimum size=0.1cm]  at (4,-1.5)  {};
\node  [draw=black,thick,circle,minimum size=0.1cm]  at (4,-1)  {};
\draw [middlearrow={latex reversed}][ultra thick,black] (7.5,-1) to[out=155,in=0] (6.5,0)
to[out=180,in=-130] (6.5,-2.4) to[out=50,in=180] (7.35,-1.5);
\node  [draw=black,fill=black,thick,circle, minimum size=0.1cm]  at (7.5,-1)  {};
\node [draw=black,thick,circle, minimum size=0.1cm]  at (7.5,-1.5)  {};
\node (5) [draw=black,thick,rotate=90,dashed,ellipse,minimum height=20pt,minimum width=35pt] 
at (7.5,-1.25) {};
\node at (3.4,-1.25) {0};
\node at (8.1,-1.25) {$x$};
\end{tikzpicture}
\caption{Connected (left) and disconnected (right) contributions to  meson
correlators. }\label{fig:CD}
\end{figure*}
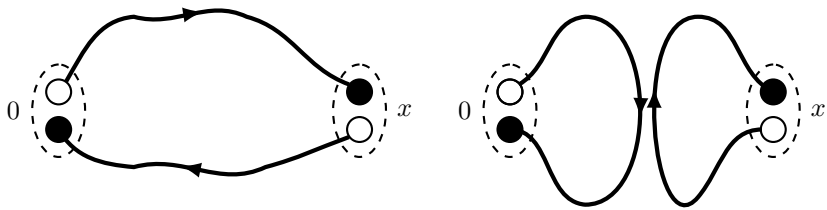

\begin{figure*}[t]
  \hspace*{12pt}
\includegraphics[scale=0.55]{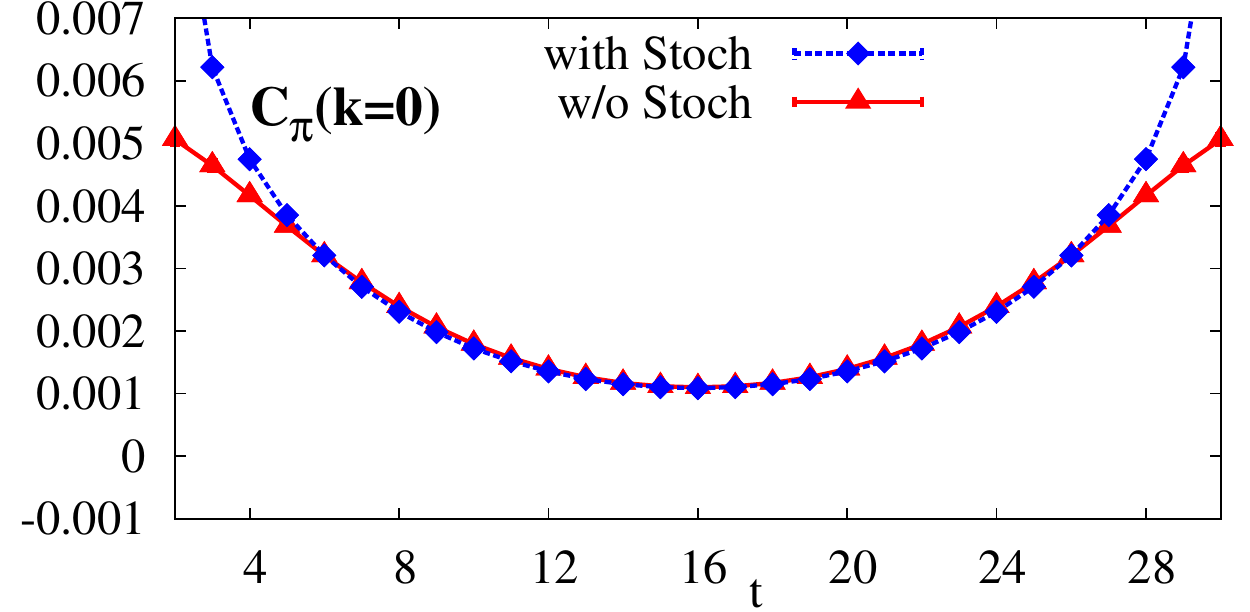} 
\includegraphics[scale=0.55]{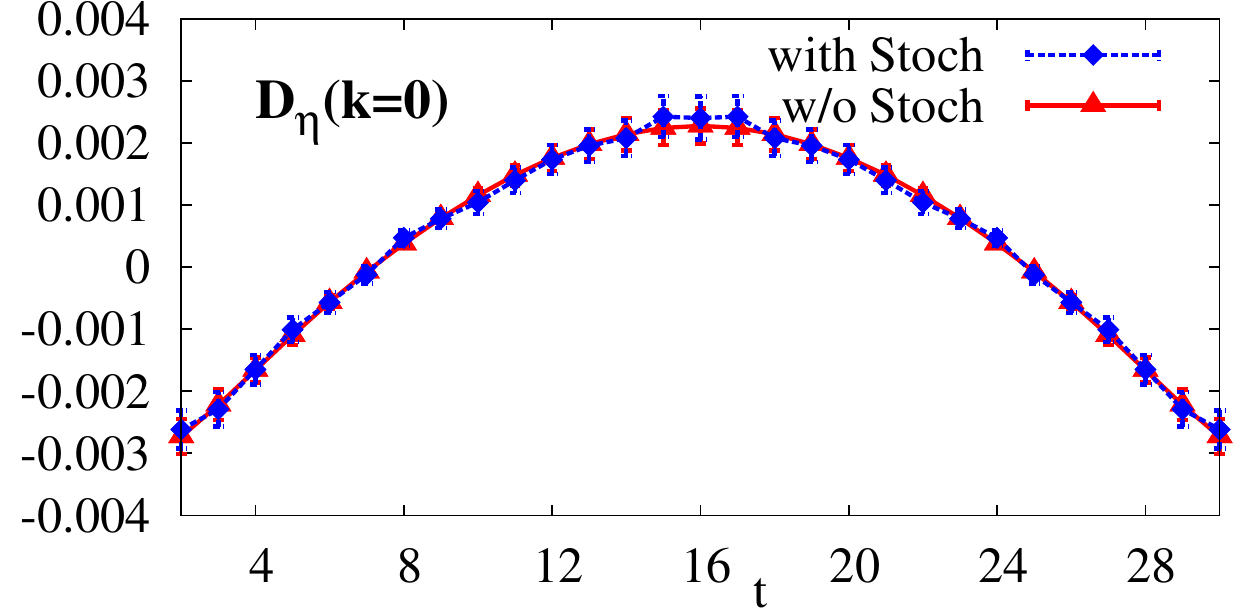} \hspace*{24pt} \hfill\\
  \hspace*{12pt}
\includegraphics[scale=0.55]{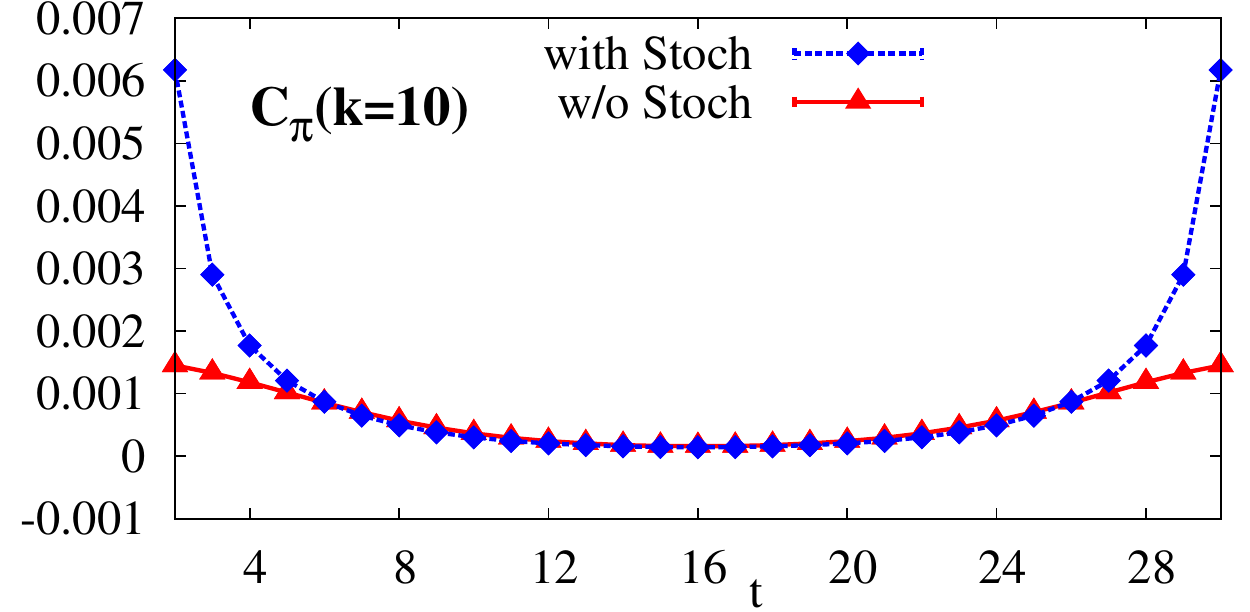}
\includegraphics[scale=0.55]{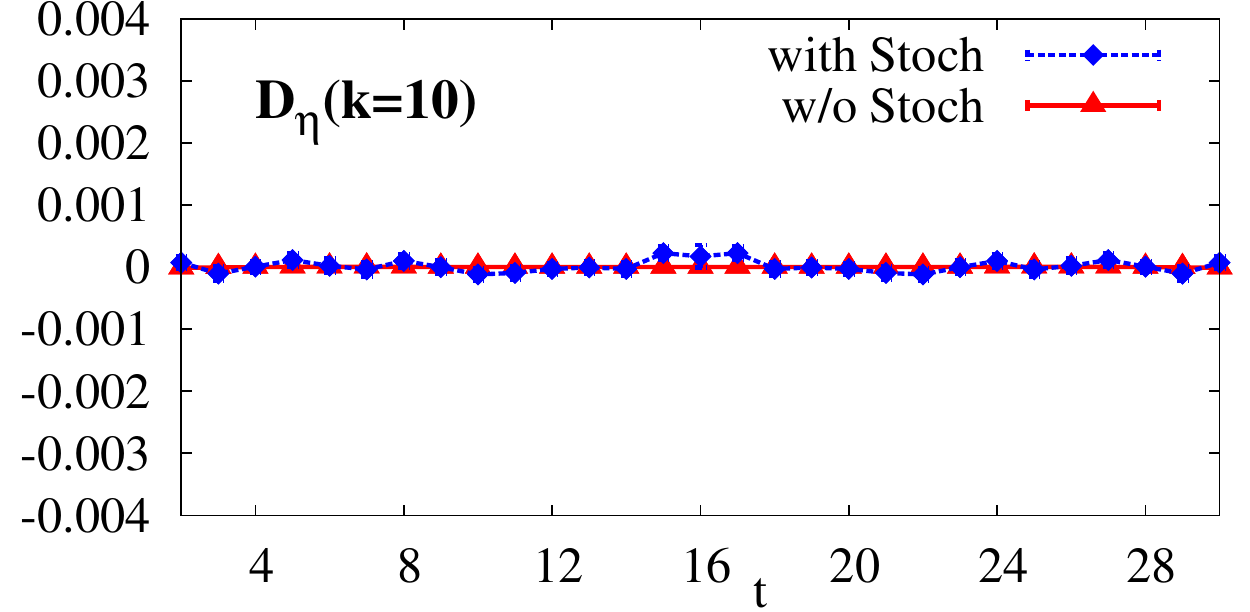}\hspace*{24pt} \hfill\\
  \hspace*{12pt}
\includegraphics[scale=0.55]{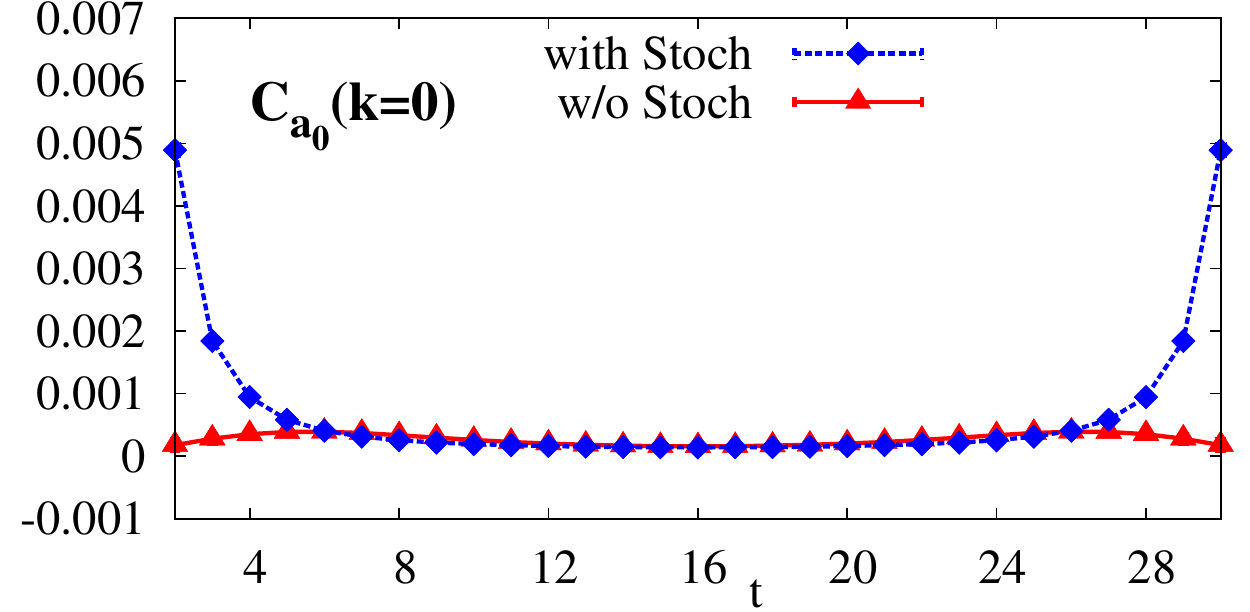} 
\includegraphics[scale=0.55]{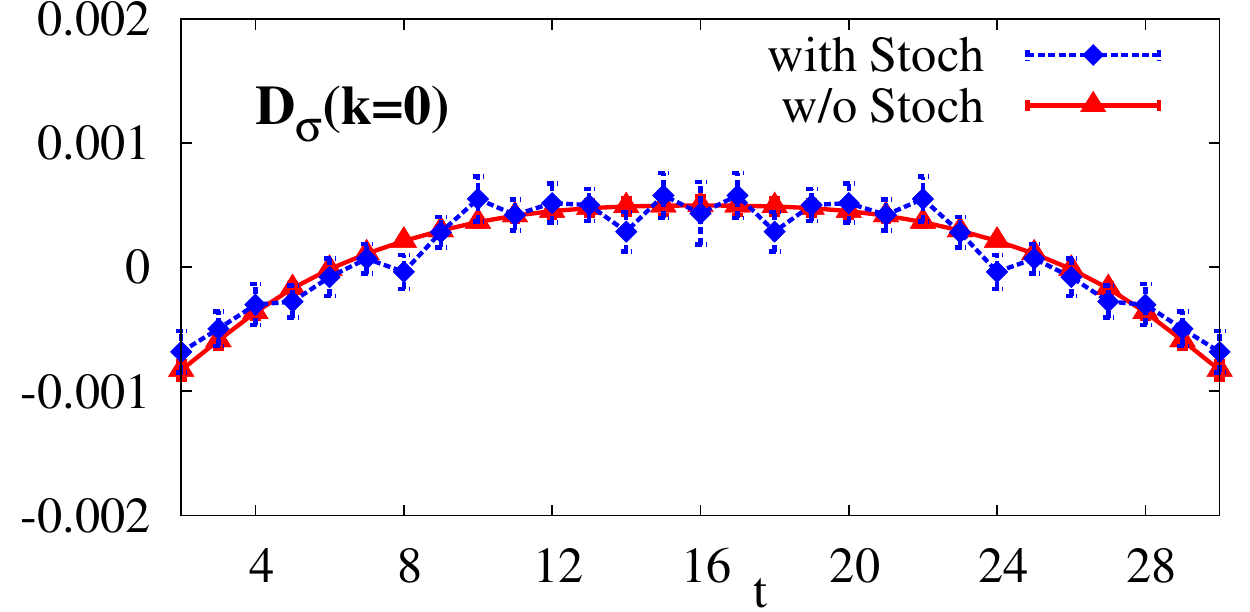} \hspace*{24pt} \hfill\\
  \hspace*{12pt}
\includegraphics[scale=0.55]{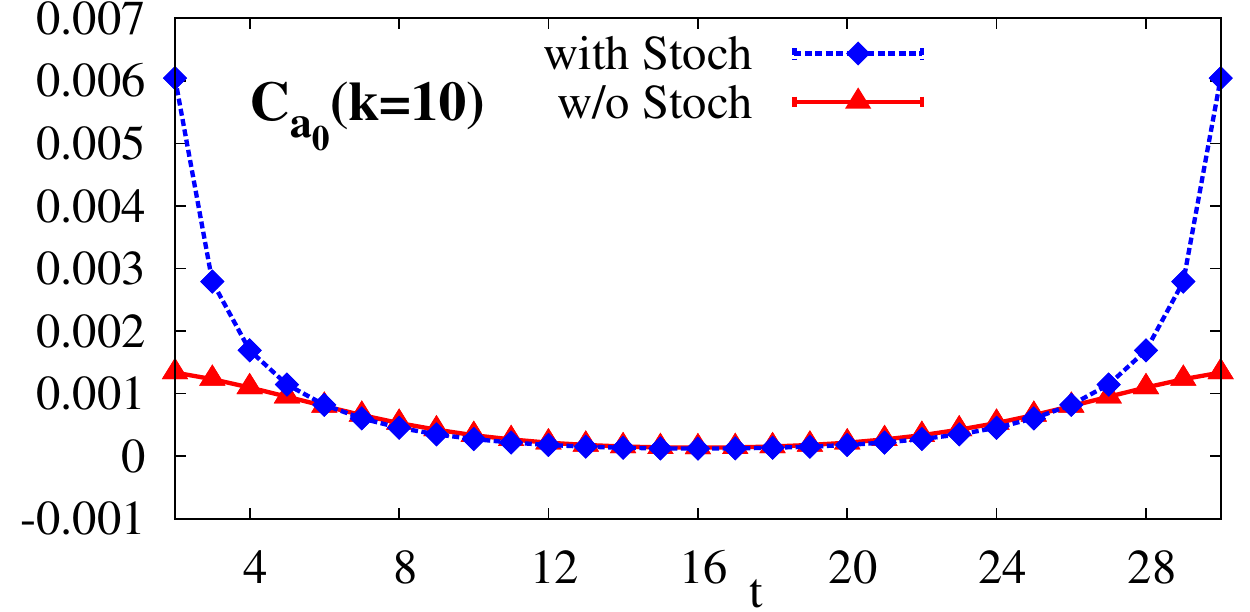}
\includegraphics[scale=0.55]{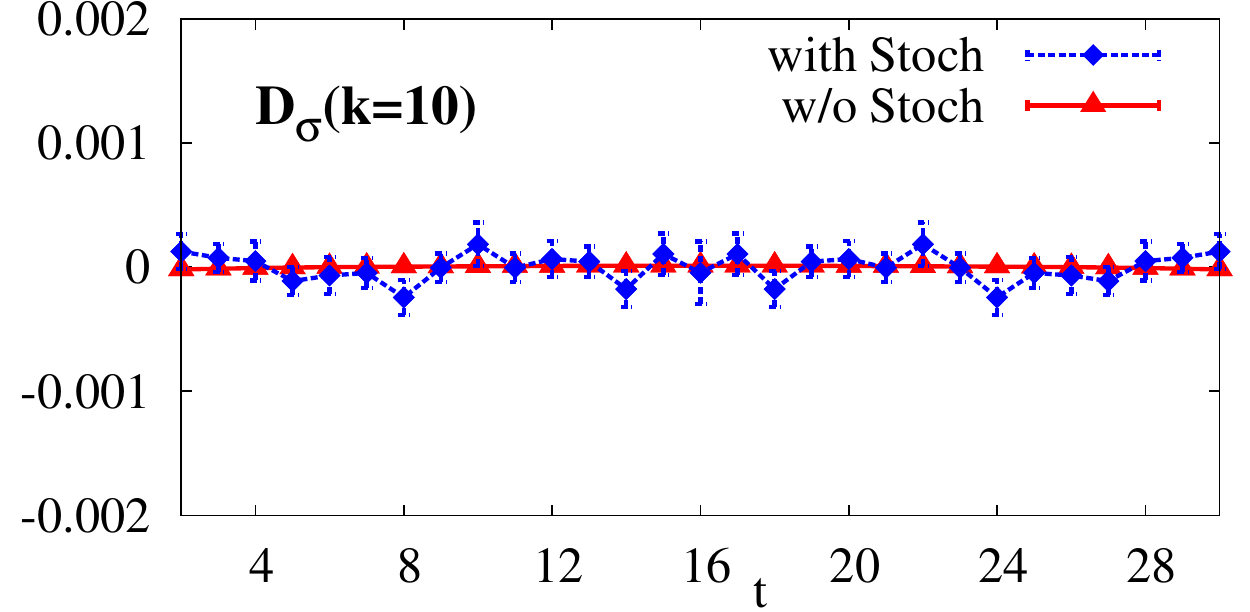}\hspace*{24pt} \hfill\\
 \caption{Connected (left) and disconnected (right) contributions to the $\eta$ and $\sigma$  meson correlator s
with included or excluded stochastically estimated part of the quark propagator,
$k=0,10$.}\label{fig:CD1}
\end{figure*}

\begin{figure*}[ht]
 \hspace*{12pt}
\includegraphics[scale=0.7]{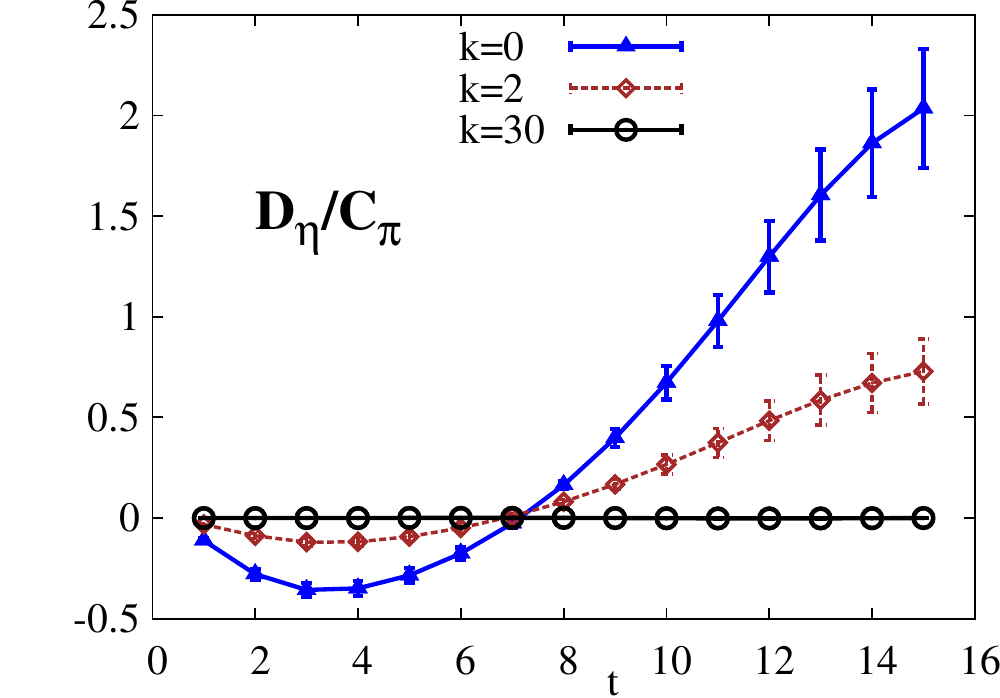}
\includegraphics[scale=0.7]{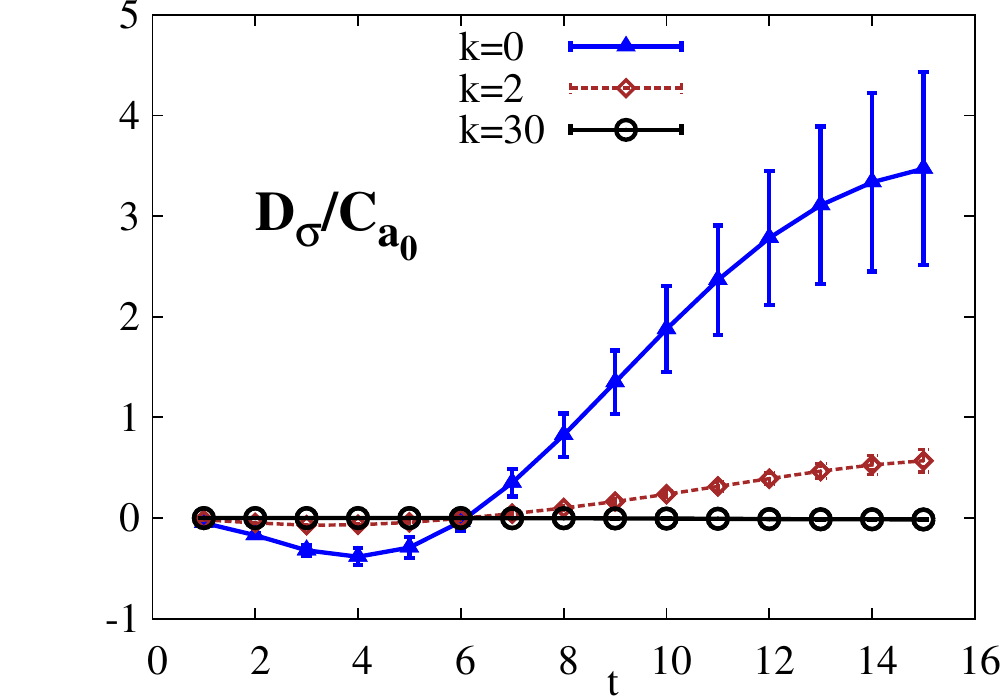}\hspace*{24pt} \hfill\\
\caption{Ratios of disconnected (without stochastic contribution) and connected $J=0$ correlators,
$k=0,2,30$.}\label{fig:ratcd1}
\end{figure*}

\begin{figure*}[t!]
  \hspace*{12pt}
  \includegraphics[scale=0.58]{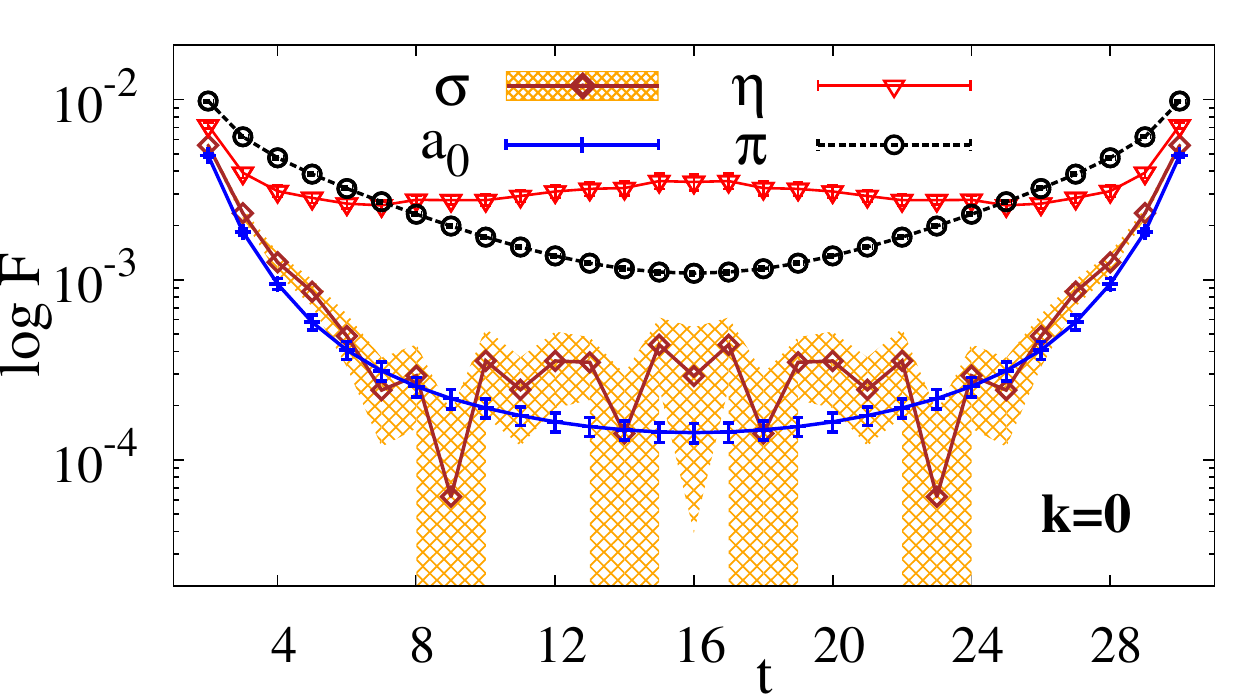}
  \includegraphics[scale=0.58]{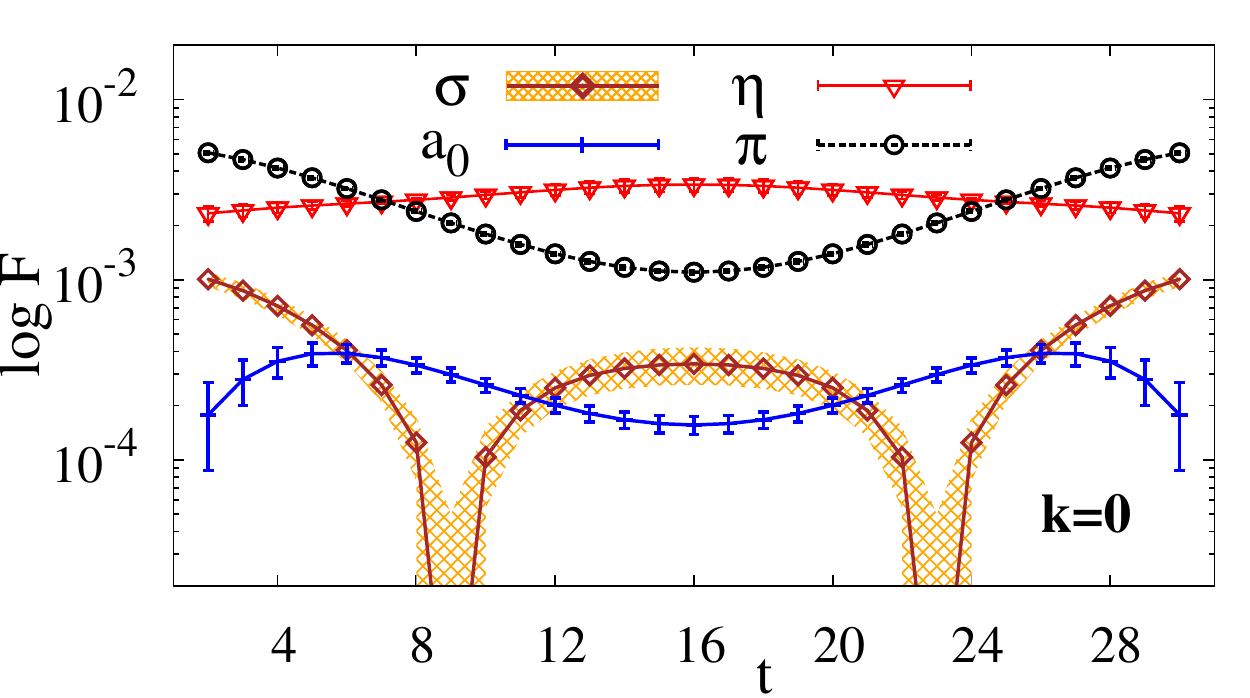}\hspace*{24pt}\hfill\\
   \caption{$\pi$, $\sigma$, $a_0$, $\eta$  correlators with stochastic contributions (left),
without stochastic contributions (right) without exclusion of the near-zero modes, $k=0$.
}
\label{fig:mescorrk0} 
  \end{figure*}
\begin{figure*}[t!]
  \hspace*{12pt}
  \includegraphics[scale=0.58]{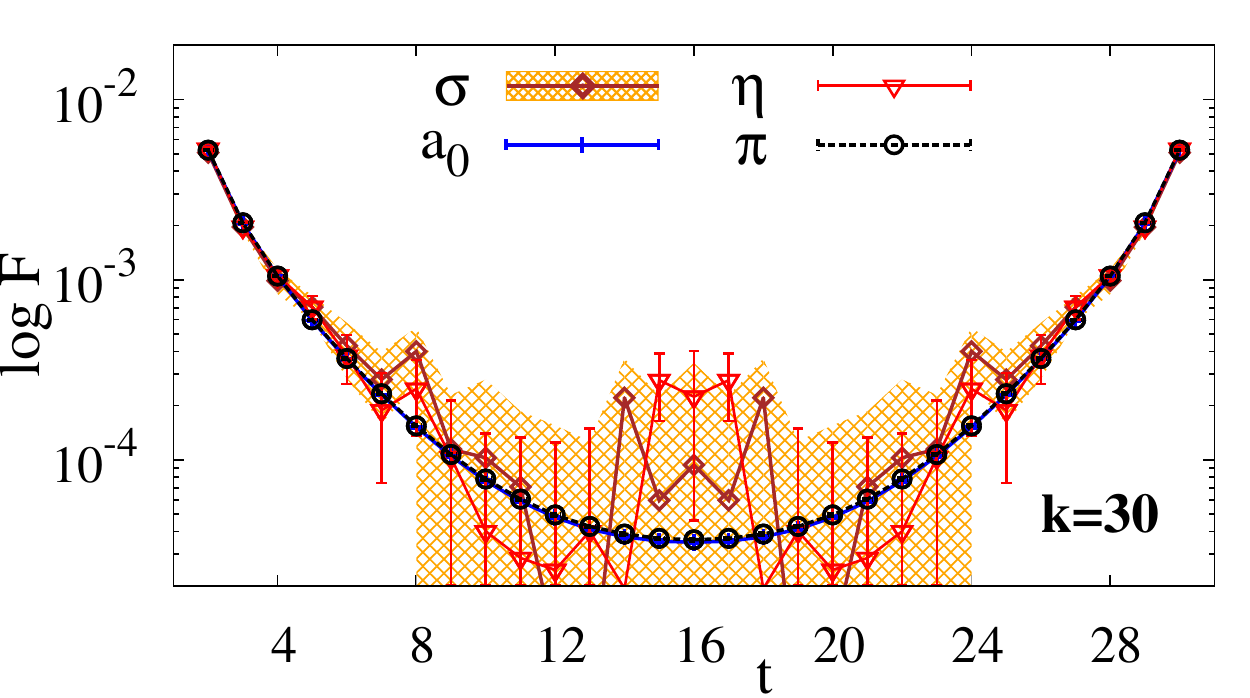}
  \includegraphics[scale=0.58]{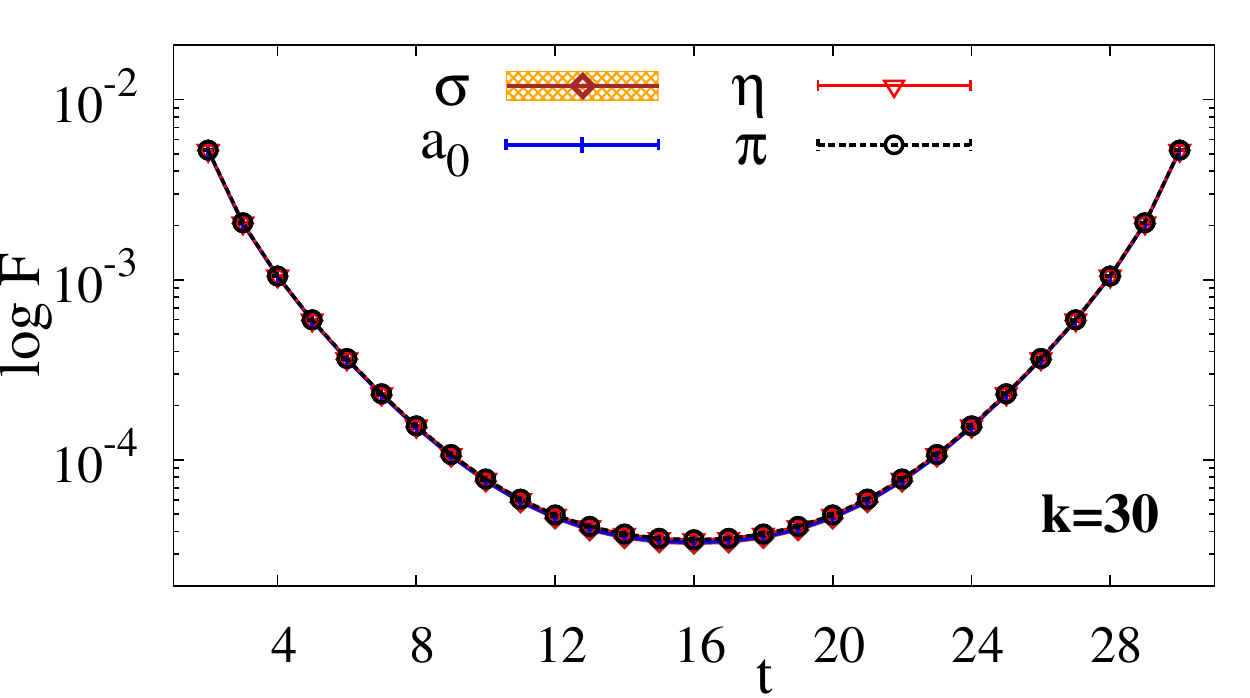}\hspace*{24pt}\hfill\\
   \caption{$\pi$, $\sigma$, $a_0$, $\eta$  correlators with stochastic contributions  (left),
without stochastic contributions to disconnected pieces (right) upon exclusion of the near-zero
modes, $k=30$. }
\label{fig:mescorrk30} 
  \end{figure*}

 \begin{figure*}[t!]
   \centering 
     \hspace*{12pt}
  \includegraphics[scale=0.53]{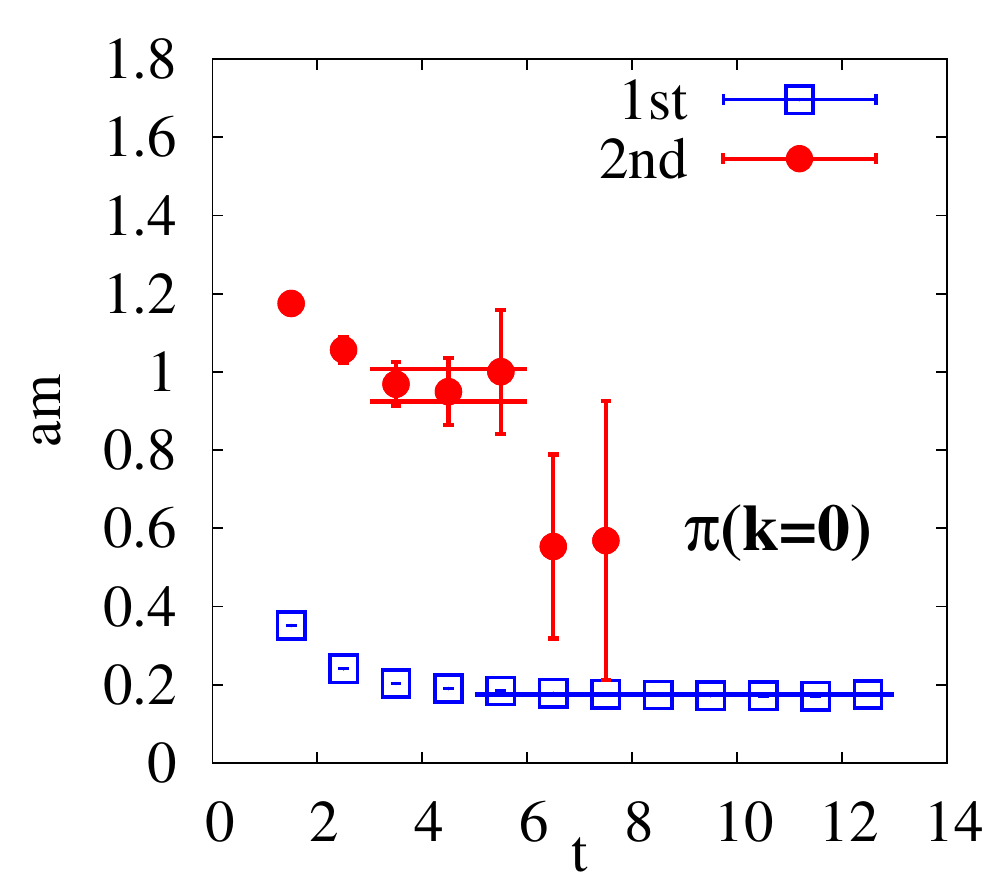}
  \hspace*{-12pt}
  \includegraphics[scale=0.53]{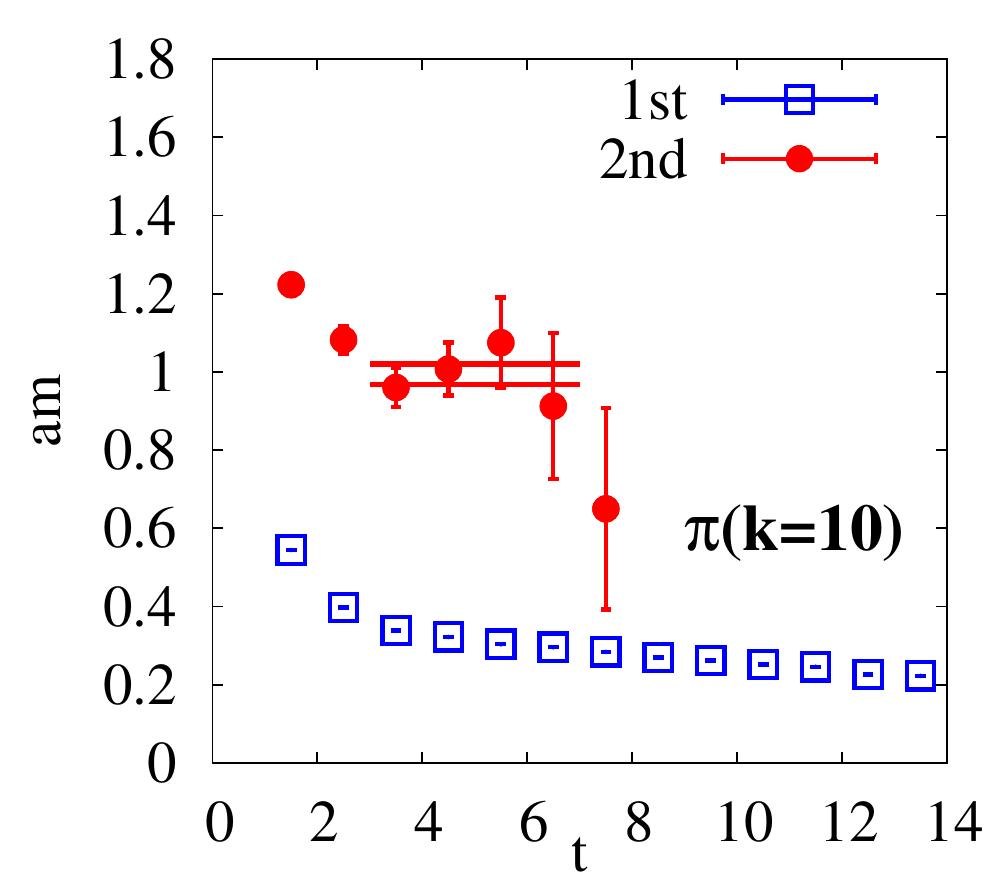}
  \hspace*{-12pt}
 \includegraphics[scale=0.53]{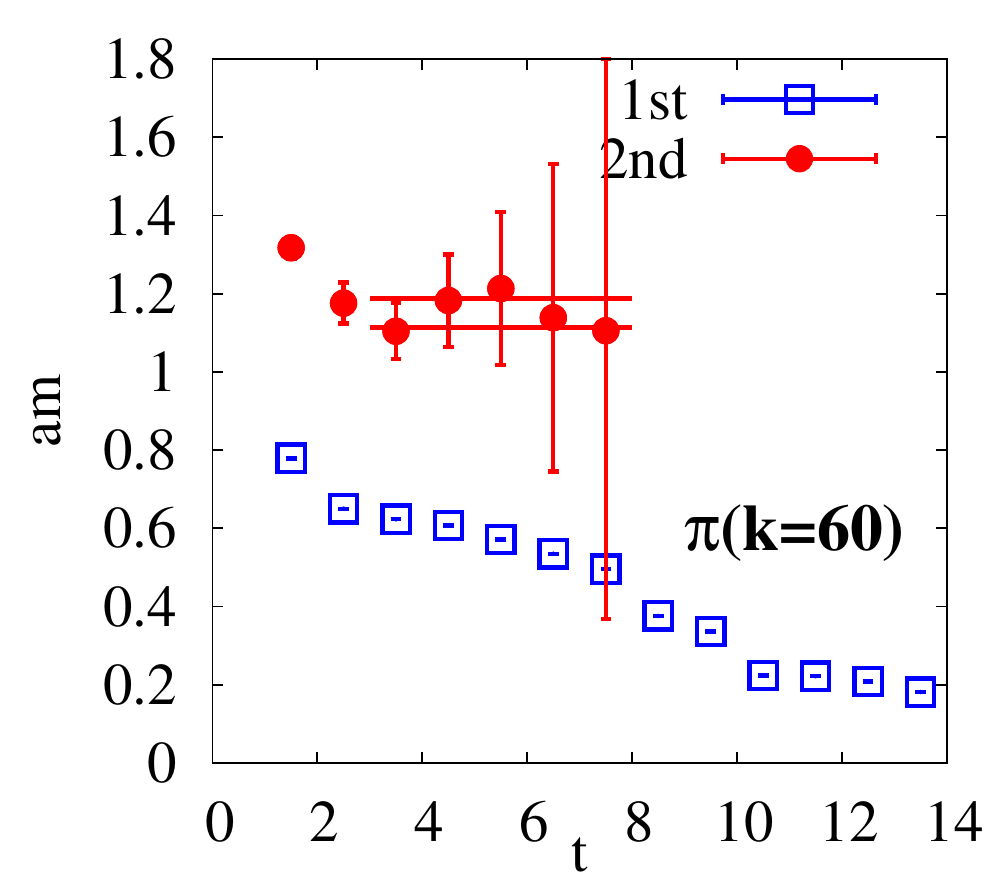}\hspace*{24pt}\hfill\\
    \caption{$\pi$ effective masses at $k=0,10,60$. }\label{fig:pimass}
\end{figure*}

 \begin{figure*}[t]
  \centering
  \hspace*{12pt}
 \includegraphics[scale=0.58]{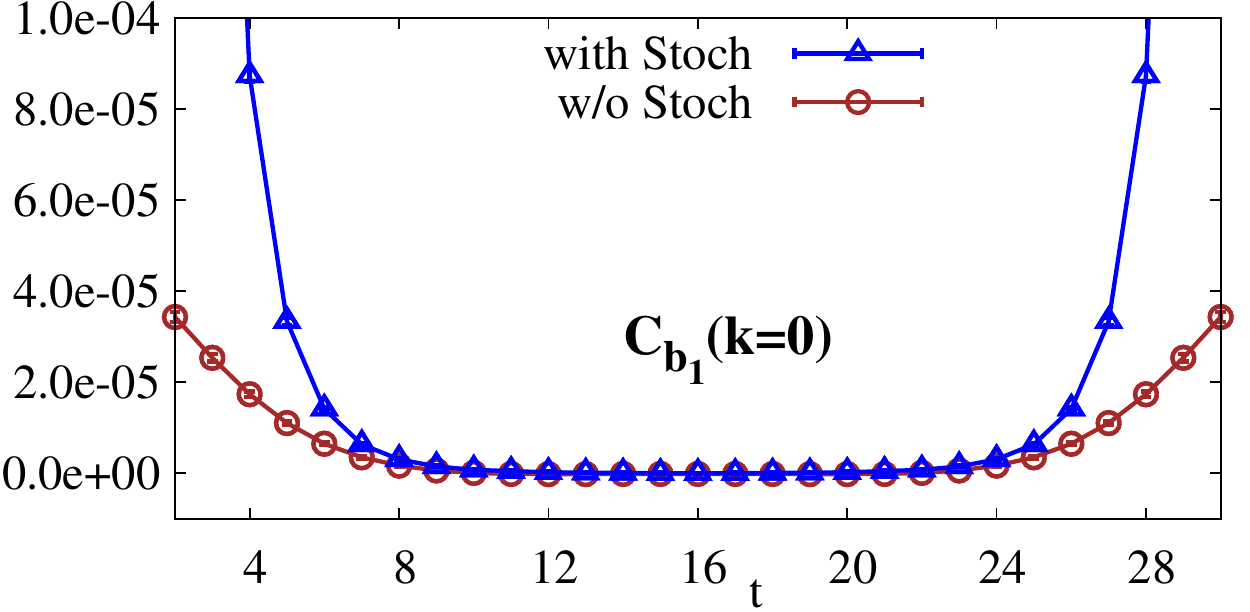}  
\includegraphics[scale=0.58]{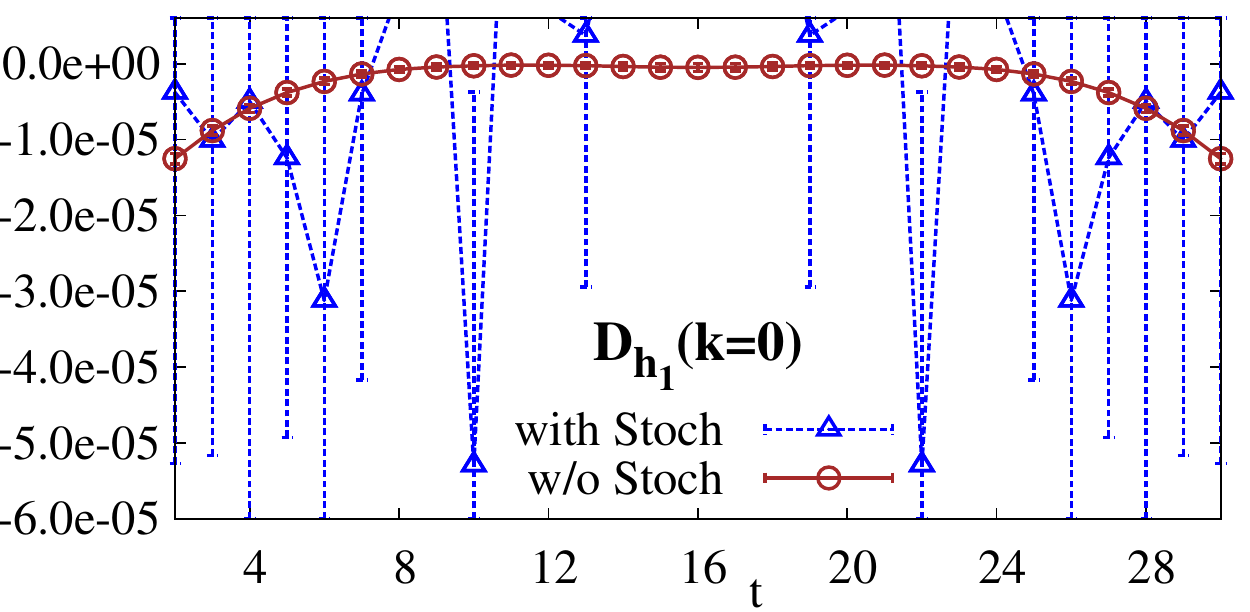} \hspace*{24pt} \hfill\\
  \hspace*{12pt}
 \includegraphics[scale=0.58]{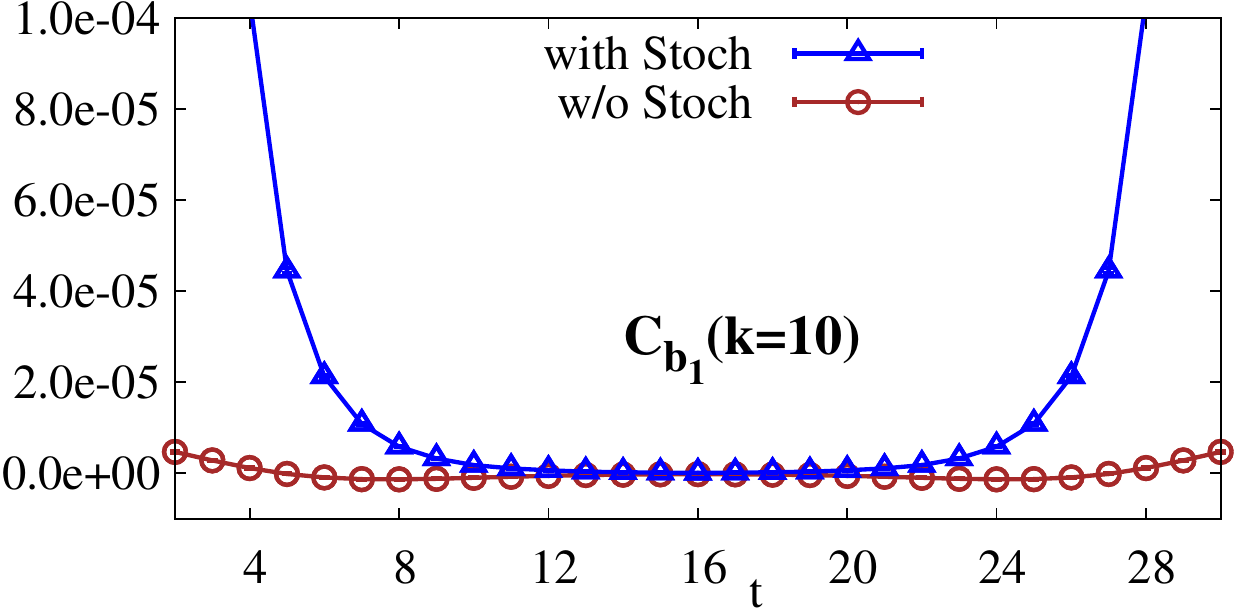}
\includegraphics[scale=0.58]{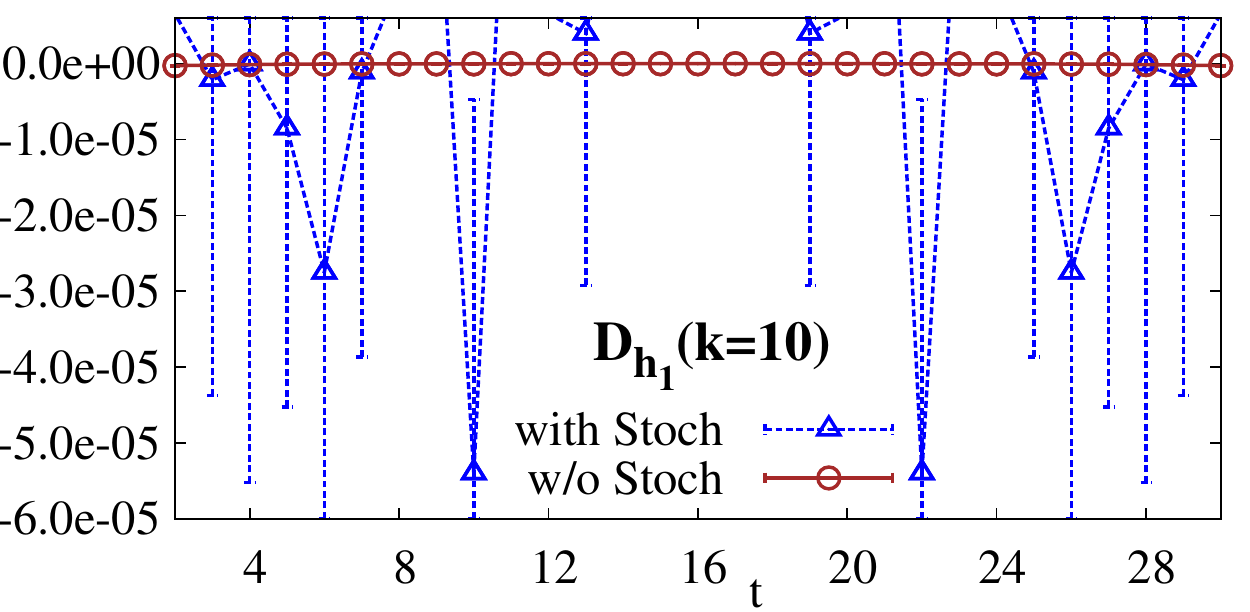}\hspace*{24pt} \hfill\\
 \caption{Connected (left) and disconnected (right) contributions to $h_1$ meson correlator 
with included or excluded stochastically estimated part of the quark propagator,
$k=0,10$.}\label{fig:CD2}
\end{figure*}

In general there are two distinct contributions to the correlation
functions depending on the isospin $I$   \cite{Kunihiro:2008dc,Hashimoto:2008xg}.
The isovector correlation functions contain only connected ($C$) contributions. Both the
connected and disconnected ($D$) diagrams contribute to the isoscalar mesons, see Fig. \ref{fig:CD}.
The difference between the isoscalar and isovector channels with the same parity, i.e.,
between $\eta$ and $\pi$ , as well as between $\sigma$ and $a_0$ is described 
by the contribution $D(t)$ coming from the disconnected dia\-grams:
\begin{equation}
F_{\eta(\sigma)}=C_{\pi(a_0)}+D_{\eta(\sigma)},
\end{equation}
where $F$ is the full correlator.

A subtraction of the quasi-zero modes of the Dirac operator should naturally
lead to the restoration of the chiral $SU(2)_L \times SU(2)_R$ symmetry since
these modes are connected
to the quark condensate of the vacuum. Consequently, upon
unbreaking of the chiral symmetry one  expects a coincidence of the $\pi$ and $\sigma$
as well as of the  $a_0$ and $\eta$ correlators. However, one does not know apriori what
will happen with the $U(1)_A$ symmetry. If the chiral symmetry is restored but the $U(1)_A$
symmetry is still broken the $\pi$ and $\eta$ correlators should be different,
i.e., the $\eta$ disconnected contributions should not vanish. If, on the contrary, the same
low-lying modes are responsible for both $SU(2)_L \times SU(2)_R$ and $U(1)_A$ breaking,
one expects that the $\eta$ disconnected contributions should vanish upon elimination of the
lowest-lying Dirac modes.

Hence, it is instructive to study the behavior of the connected and disconnected correlators  upon
exclusion of the near-to-zero modes separately from each other. In  
Fig. \ref{fig:CD1} we show the connected ($\pi$) and disconnected ($\eta$) contributions
for untruncated correlators ($k=0$) and after subtraction of 10 lowest eigenmodes of the
Dirac operator ($k=10$). It is clearly seen that in the untruncated case both the connected
and disconnected contributions are equally important, which provides an essential difference
of the $\pi$ and $\eta$ masses.  Actually, for the $\eta$ propagator they almost cancel, prohibiting
a determination of the mass. This is due to the fixing of the global topological charge which leads to a vanishing
topological susceptibility $\chi_T$ (In Ref. \cite{Aoki_2008_topo} it is discussed how to 
estimate $\chi_T$ in this case from the disconnected part of the $\eta$
correlator).

In the figures we exhibit both, the propagators with and without the stochastic contributions to
the quark propagators (see Eq. (\ref{def_trunc})).
We see, that the stochastic part is very small beyond $t>5$ and negligible for the disconnected parts.
The dependence of the disconnected parts with truncation is remarkable.
Upon elimination of 10 lowest Dirac eigenmodes
the disconnected contribution practically vanishes. This signals a simultaneous 
restoration of both symmetries. 

The same observation holds for the connected  and disconnected contributions
in the $a_0$ and $\sigma$ channels. The ratios of the corresponding contributions are
shown in Fig. \ref{fig:ratcd1}.
We conclude that the
same lowest-lying eigenmodes of the Dirac operator are responsible for both 
$SU(2)_L \times SU(2)_R$ and $U(1)_A$ breakings which is consistent with the instanton-induced
mechanism of both breakings \cite{H,S,D}. 

The total ($F$)  point-to-point correlators in all $\pi,\sigma,a_0,\eta$ channels are shown in
Fig. \ref{fig:mescorrk0}-\ref{fig:mescorrk30}. For the untruncated case we see again by comparing the plots in Fig. \ref{fig:mescorrk0} left to right, that the stochastic contribution is relevant for small distances and 
responsible for the increased noise in the $\sigma$ correlation. For the truncated case (Fig. \ref{fig:mescorrk30}) 
the results agree whether one includes the stochastic contribution or not. An exception is again the
 $\sigma$ where the noise due to the stochastic tern shadows the large distance behavior. 
The unusual shape of  the $\eta$
correlator  is due to the fixing of the global topological as discussed in Ref.  \cite{Aoki:2012pma}.
Without truncation of the lowest-lying Dirac modes the correlators  are all very different due to 
breakings of both  $SU(2)_L \times SU(2)_R$ and $U(1)_A$  symmetries. After truncation of a small amount 
of the quasi-zero modes they become all almost identical. This tells once again that both symmetries are restored.  

The next natural question is to ask whether the states still exist as physical states in this chirally restored
regime. To answer this question we concentrate on the pion channel for the following  reason: The
original $\pi$  states can be easily identified in the untruncated case. 
Extraction of good effective mass plateaus in other quantum channels requires much
better statistics.

 \begin{figure*}[ht]
  \centering
  \hspace*{8pt}
\includegraphics[scale=0.65]{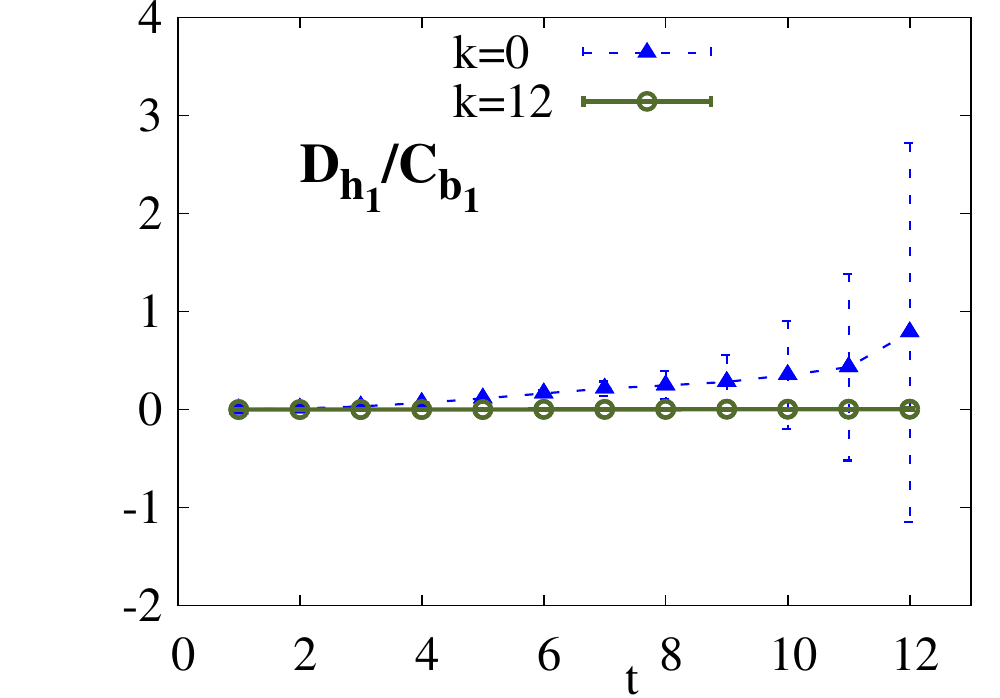} 
\includegraphics[scale=0.65]{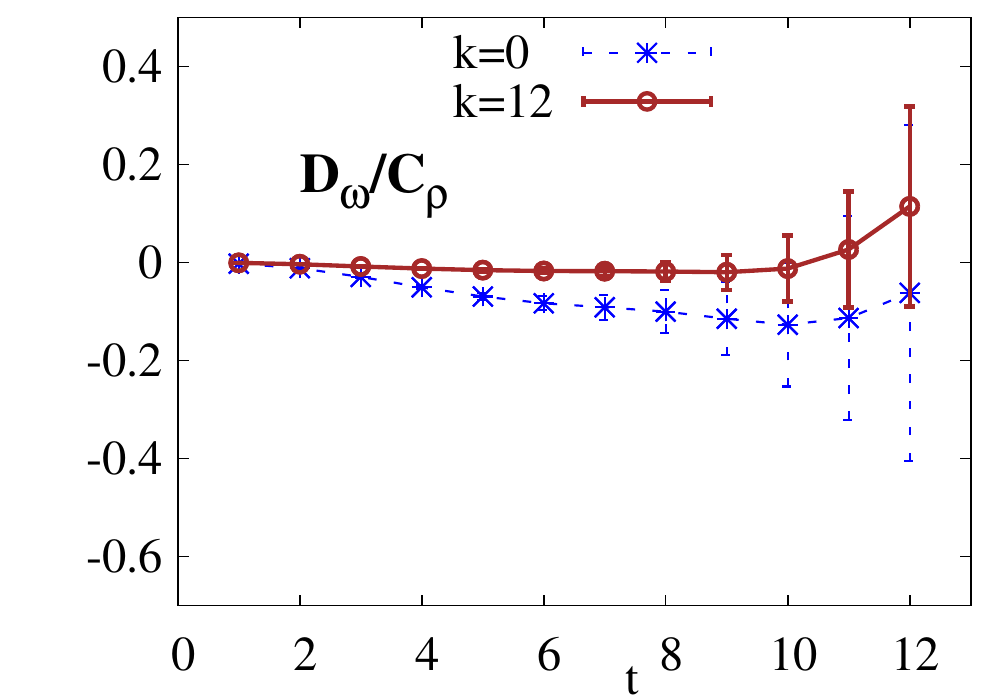}\hspace*{24pt} \hfill\\
\caption{Ratios of the disconnected and
connected $J=1$ correlators,
$k=0,12$. }\label{fig:ratcd2}
\end{figure*} 

\begin{figure*}[ht]
  \hspace*{12pt}
  \includegraphics[scale=0.58]{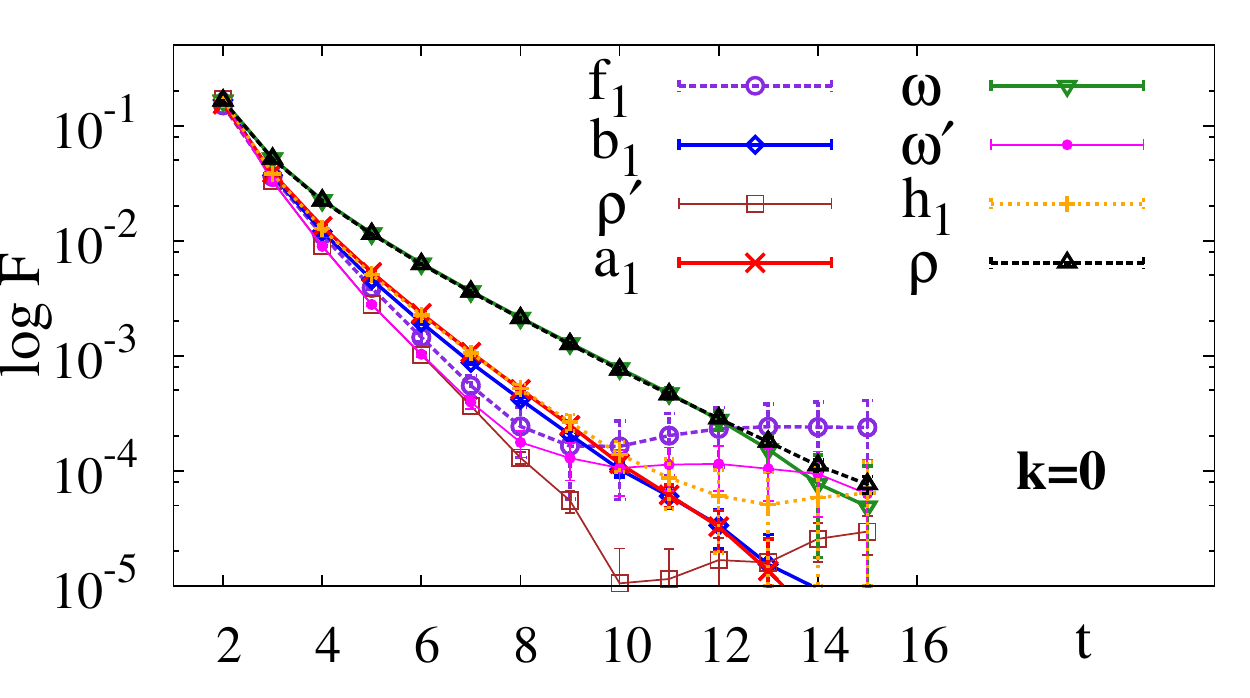}
  \includegraphics[scale=0.58]{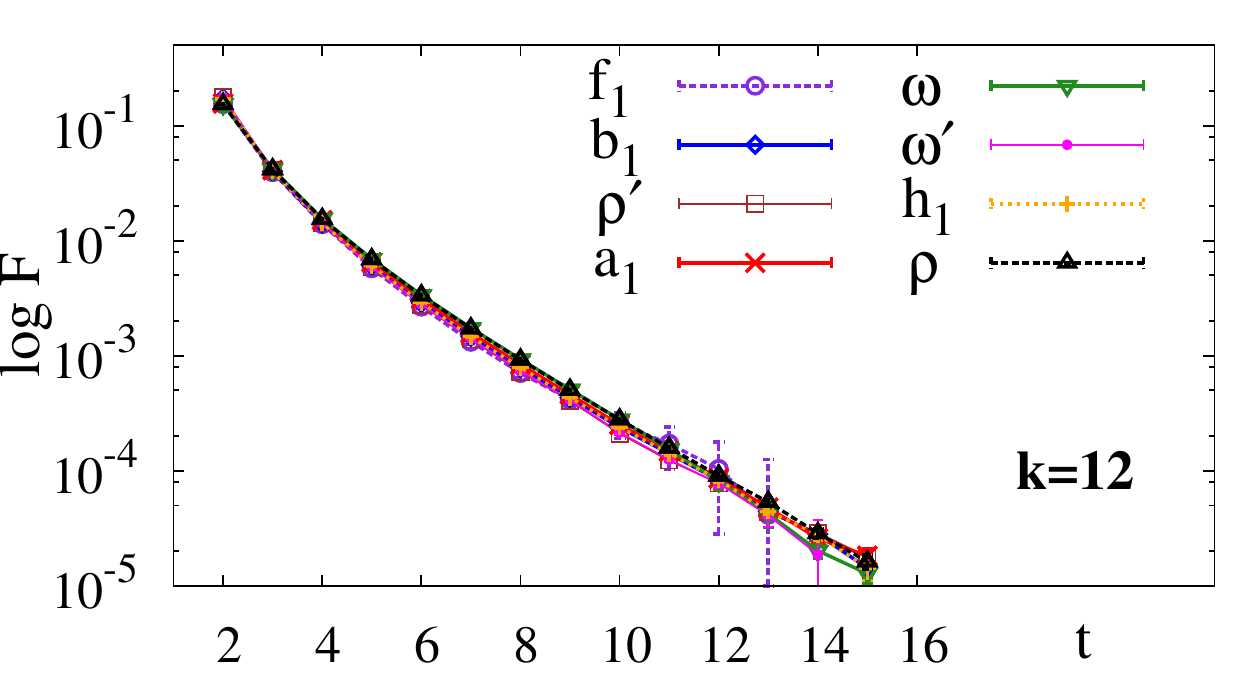}\hspace*{24pt}\hfill\\
   \caption{ $J=1$  correlators without truncation (left) and upon exclusion of the near-zero modes,
$k=12$ (right).
}
\label{fig:evals1} 
  \end{figure*}
  
On exclusion of the near-zero modes the ground state effective mass
plateau of the pion deteriorates and disappears, see Fig. \ref{fig:pimass}. The correlation
function decays with time not exponentially and thus the eigenstate
is not a physical state. Hence unbreaking of the chiral symmetry
removes the pion  from the physical spectrum. This is consistent with the Goldstone boson nature of the pion.
The quasi-zero modes play an important role in the pion and are crucial for its existence.
The $\pi'$ state might however survive the unbreaking, though to firmly conclude it one needs better statistics.

A sufficient condition for a symmetry restoration is a coincidence of
the correlators obtained with the operators that are connected by the
symmetry transformation. From the results presented above it then follows
that the $SU(2)_L \times SU(2)_R$ and $U(1)_A$ symmetry in the $J=0$ sector
is restored upon unbreaking. Existence or nonexistence of the bound states
in the symmetry restored regime is the other question. If at a given energy at least
in one of the channels there is no bound state, then the symmetry requires
that there must not be bound states at the same energy in all other $J=0$
channels. The pion does not survive the unbreaking. Consequently there
cannot be respective $\sigma, a_0, \eta$ bound states after unbreaking.
The excited $\pi'$ state might survive, however. If it does, there must
be then degenerate $\pi',\sigma', a_0', \eta'$ states in the spectrum.

\subsection{ $J=1$ mesons}
   \begin{figure}[h]
 \begin{tikzpicture}[<->,>=stealth',shorten >=1pt,auto,font=\small,
                    thick,main
node/.style={rectangle,draw=none,text width=5cm,text centered,font=\small},]
  \node[node distance=3.5cm] (1) {$b_1$ };
  \node[node distance=3.5cm] (2) [right of=1] {$\omega^{\prime}$};
  \node[node distance=1.2cm] (3) [below of=2] {$h_1$};
  \node[node distance=1.2cm] (4) [below of=1] {$\rho^{\prime}$}; 
    \node[node distance=0.55cm] (5) [above of=1]{ $\qquad f_1 \qquad$};
       \node[node distance=3.5cm] (6) [right of=5] {$\qquad \omega\qquad $ };
    \node[node distance=0.55cm] (7) [below of=4]{$a_1$
};
       \node[node distance=3.5cm] (8) [right of=7] {$\rho$};
         \node[node distance=2cm] (9) [left of=5]{$(0,0)$:};
    \node[node distance=2cm] (10) [left of=1]{ $\;(\frac{1}{2},\frac{1}{2})_a$:}; 
     \node[node distance=2cm] (11) [left of=4]{ $\;(\frac{1}{2},\frac{1}{2})_b$:};  
      \node[node distance=2cm] (12) [left of=7]{$(1,0) + (0,1)$:};     
  \path[every node/.style={font=\sffamily}]
    (1) edge [] node[ pos=0.5, sloped, above] {\small $SU(2)_L\times SU(2)_R$}(2)
    (2) edge [] node[ pos=0.5,  right] {\small $U(1)_A$}                       (3)
    (3) edge [] node[ pos=0.5, sloped, below] {\small $SU(2)_L\times SU(2)_R$}(4)
    (4) edge [] node[ pos=0.5, left] {\small $U(1)_A$}                         (1)
    (7) edge [] node[ pos=0.5, sloped, below] {}(8)    ; 
\end{tikzpicture}
   \caption{Symmetry relations among $J=1$ mesons \cite{Glozman:2007ek,G3}.}\label{fig:symj1}
   \end{figure}
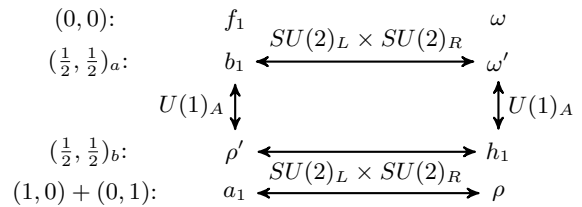
   
    \begin{figure*}[t!]
   \centering 
     \hspace*{12pt}
 \includegraphics[scale=0.61]{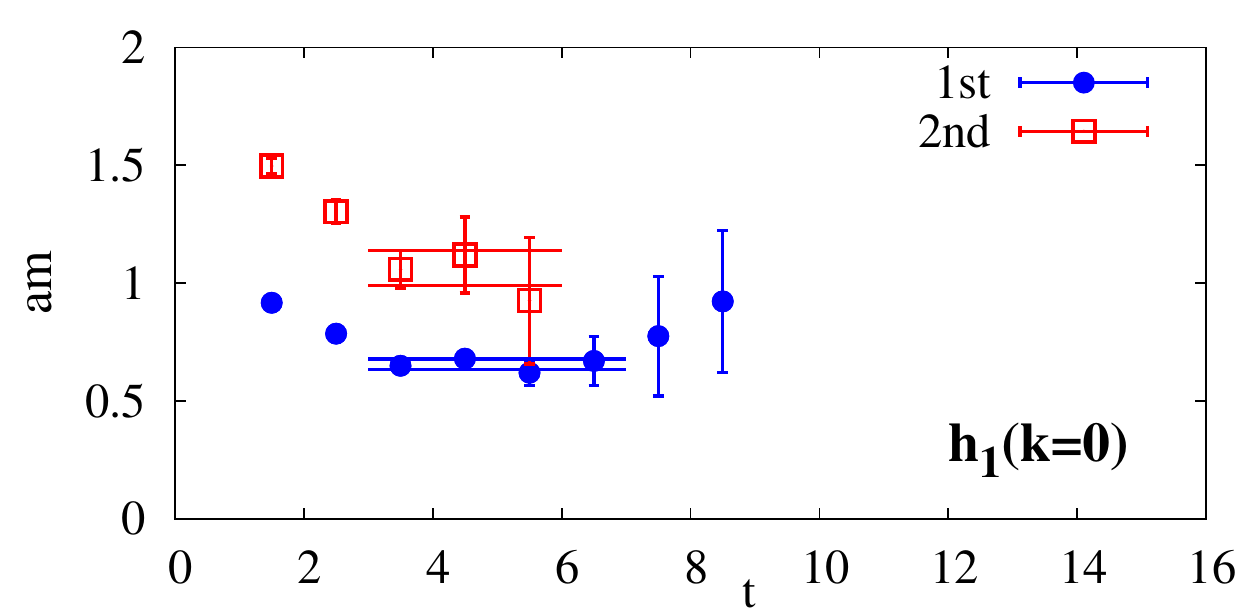}
 \includegraphics[scale=0.61]{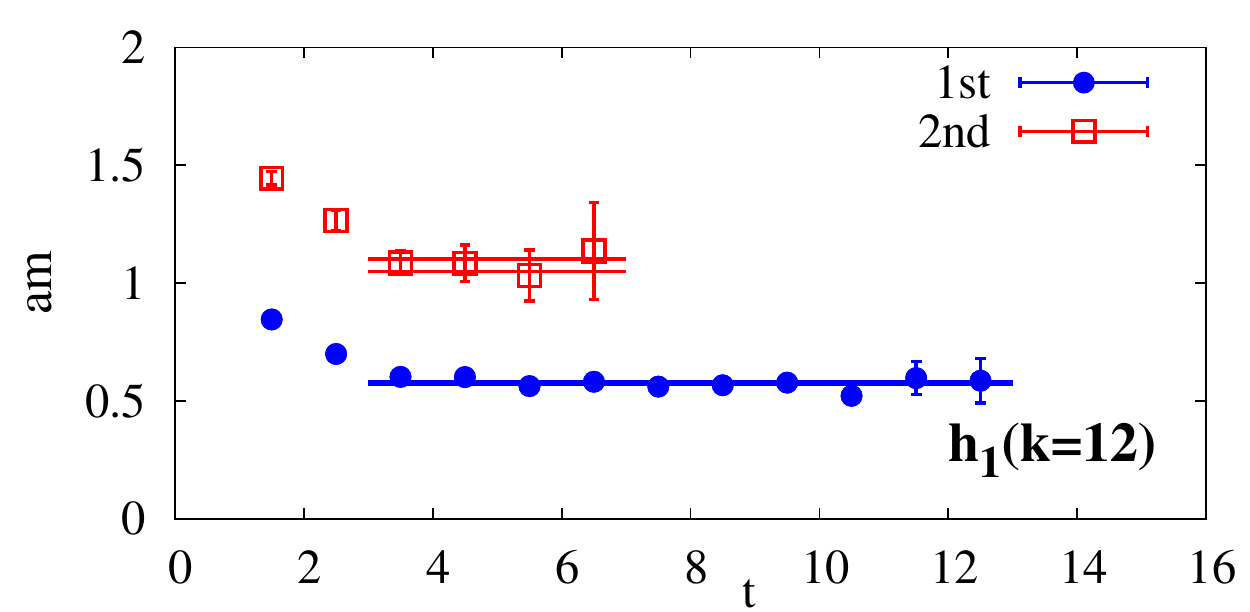}\hspace*{24pt}\hfill\\
  \hspace*{12pt}
 \includegraphics[scale=0.61]{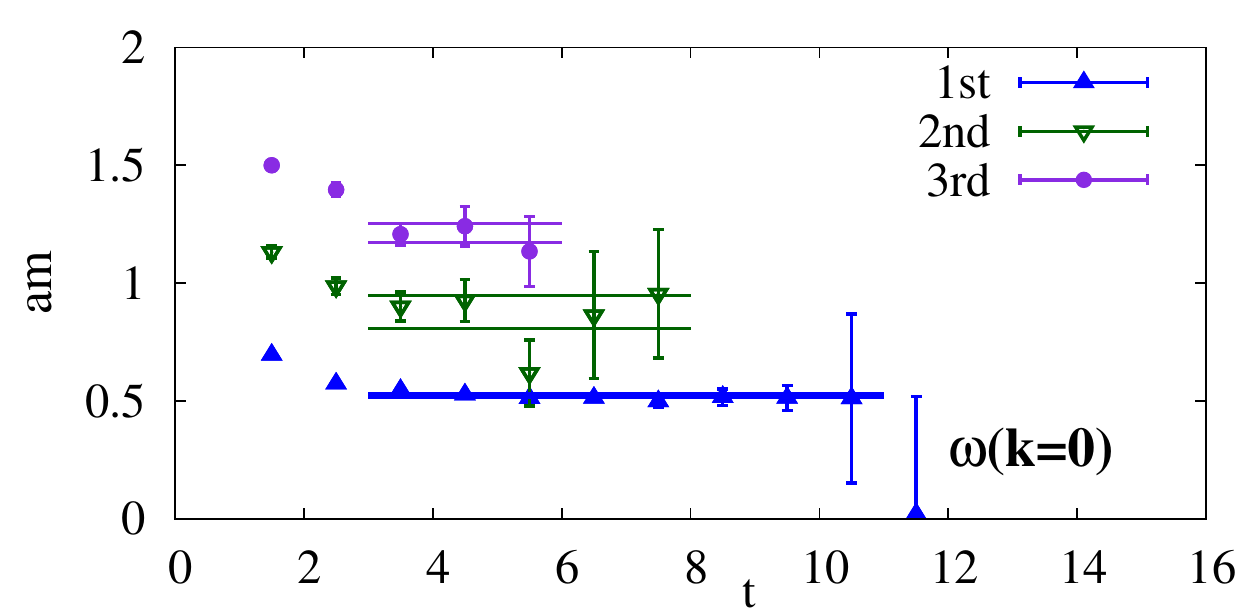}
 \includegraphics[scale=0.61]{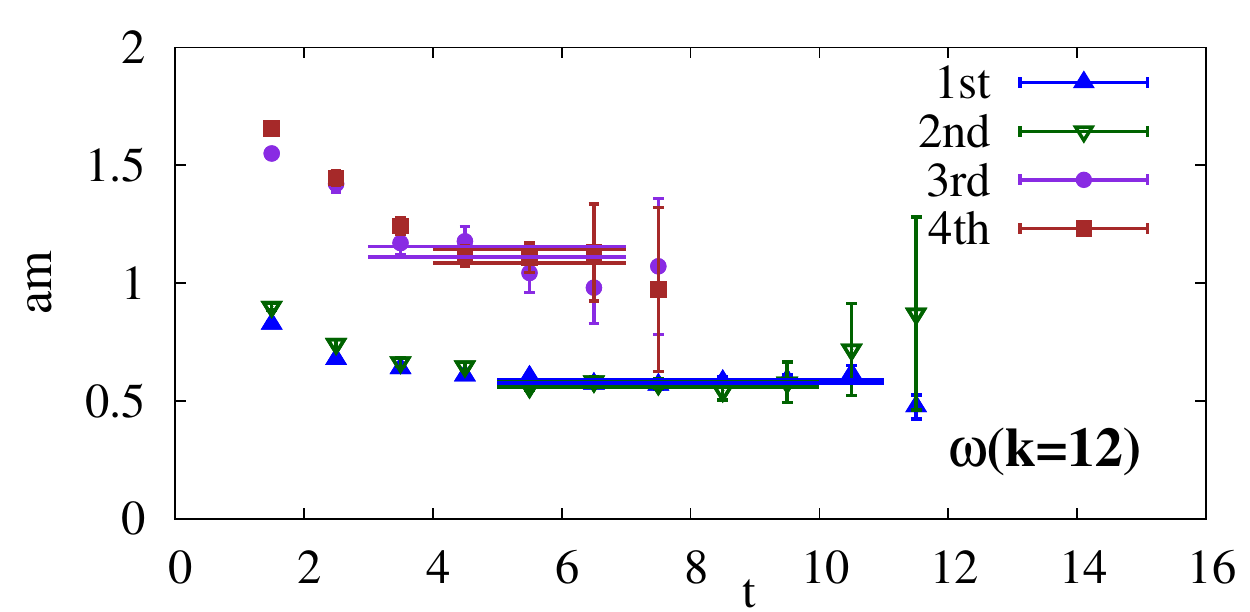}\hspace*{24pt}\hfill\\
   \hspace*{12pt}
  \includegraphics[scale=0.61]{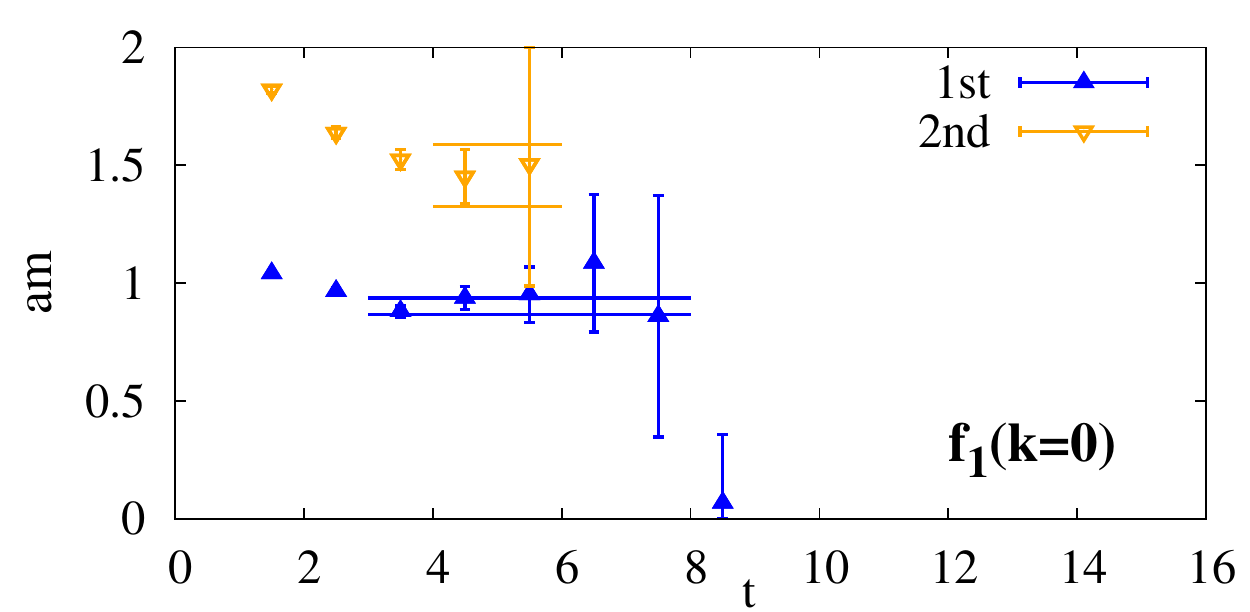}
 \includegraphics[scale=0.61]{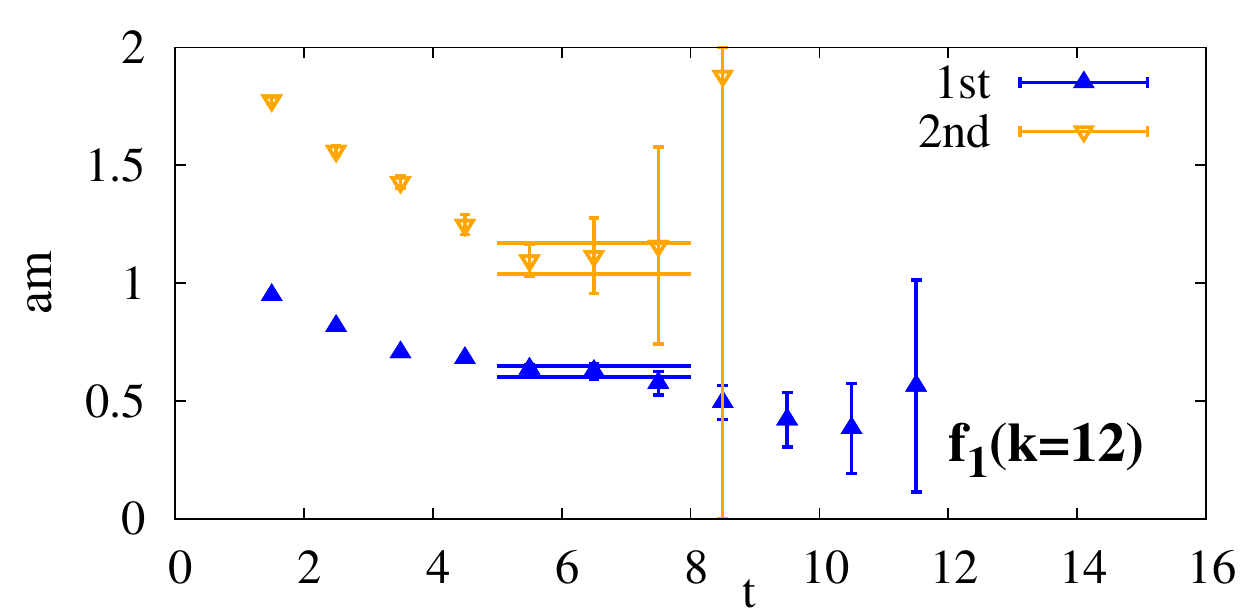}\hspace*{24pt}\hfill\\ 
    \caption{Effective masses for $h_1, \omega, f_1$ isoscalars. }\label{fig:j1mass}
  \end{figure*} 
  
   Like the isoscalar $J=0$ mesons, the isoscalar $J=1$ mesons also
contain both connected  and disconnected contributions. In contrast to the $J=0$ channels, the
disconnected contributions in the $J=1$ case are small compared to the connected
 ones. Indeed, the ground states of $\rho$ and $\omega$
 mesons are approximately degenerate in Nature. Microscopically their splitting is due
 to contributions of the disconnected graphs. Hence the latter contributions have to be
 small compared to the connected ones. The most likely interpretation is that these
 disconnected diagrams are supported in the infrared by the instantons, i.e., by the 't Hooft
 vertex. This vertex is limited, however, to the $J=0$ channel.  Consequently, while for  the $J=0$
mesons a removal of the lowest Dirac eigenmodes from the valence quarks  diminishes the otherwise
large disconnected contributions, for the spin $J=1$ mesons we should  not expect any significant
changes in the disconnected correlators.

The disconnected contributions for $J=1$ are small, very noisy and suffer from large fluctuations,
see Fig. \ref{fig:CD2}. From the results in the $J=0$ sector we have seen that the 
contributions to the disconnected terms from the stochastic part are negligibly small. Consequently,
we  have decided to omit the small but very noisy stochastically estimated contribution of the
high-lying modes to the disconnected correlators and discuss in this case only the exactly calculated
contribution  coming from the lowest 100 modes.
 
In Fig. \ref{fig:ratcd2} we show the ratio of the disconnected to the connected contribution in the
$h_1$  correlator. The connected  $h_1$ and $b_1$
contributions coincide. We see that the
disconnected contribution is very small even for
the ``untruncated'' case $k=0$ and vanishes practically
completely upon removal of the
lowest 12 Dirac eigenmodes. A similar behavior is found in other $J=1$ isoscalar channels.

The point-to-point correlators obtained with the $J=1$ operators from the 
Table 1 are shown in Fig. \ref{fig:evals1}. The symmetry connections between these
operators are illustrated in Fig. \ref{fig:symj1}. A coincidence of the
correlators in the $\rho - a_1$, $b_1 -
\omega^\prime$ and $h_1 - \rho'$ pairs evidences a restoration of $SU(2)_L \times SU(2)_R$. The equality of the $h_1, \rho', b_1, \omega^\prime$ correlation functions demonstrates the  $SU(2)_L \times SU(2)_R \times U(1)_A$
symmetry. A coincidence of all eight correlators tells about some higher
symmetry; we will return to this question below.

Now we turn to the discussion of the bound states in all these channels. The bound states of the isovector mesons $\rho,\rho',a_1,b_1$ have been shown in our previous paper \cite{DGL}.
Fig. \ref{fig:j1mass} demonstrates effective mass plateaus for all isoscalar $J=1$ mesons upon
truncation of the lowest-lying Dirac eigenmodes. We see a reliable mass plateau
for the ground state $h_1$, which significantly improves upon truncation of the lowest eigenmodes.
Upon unbreaking of chiral symmetry the ground $\omega$-state and its first excitation
$\omega^{\prime}$ get degenerate. It is similar to what happens to $\rho$ and $\rho^{\prime}$
states \cite{DGL}.

    \begin{figure*}[t!]
    \centering 
     \hspace*{20pt}
  \includegraphics[scale=0.95]{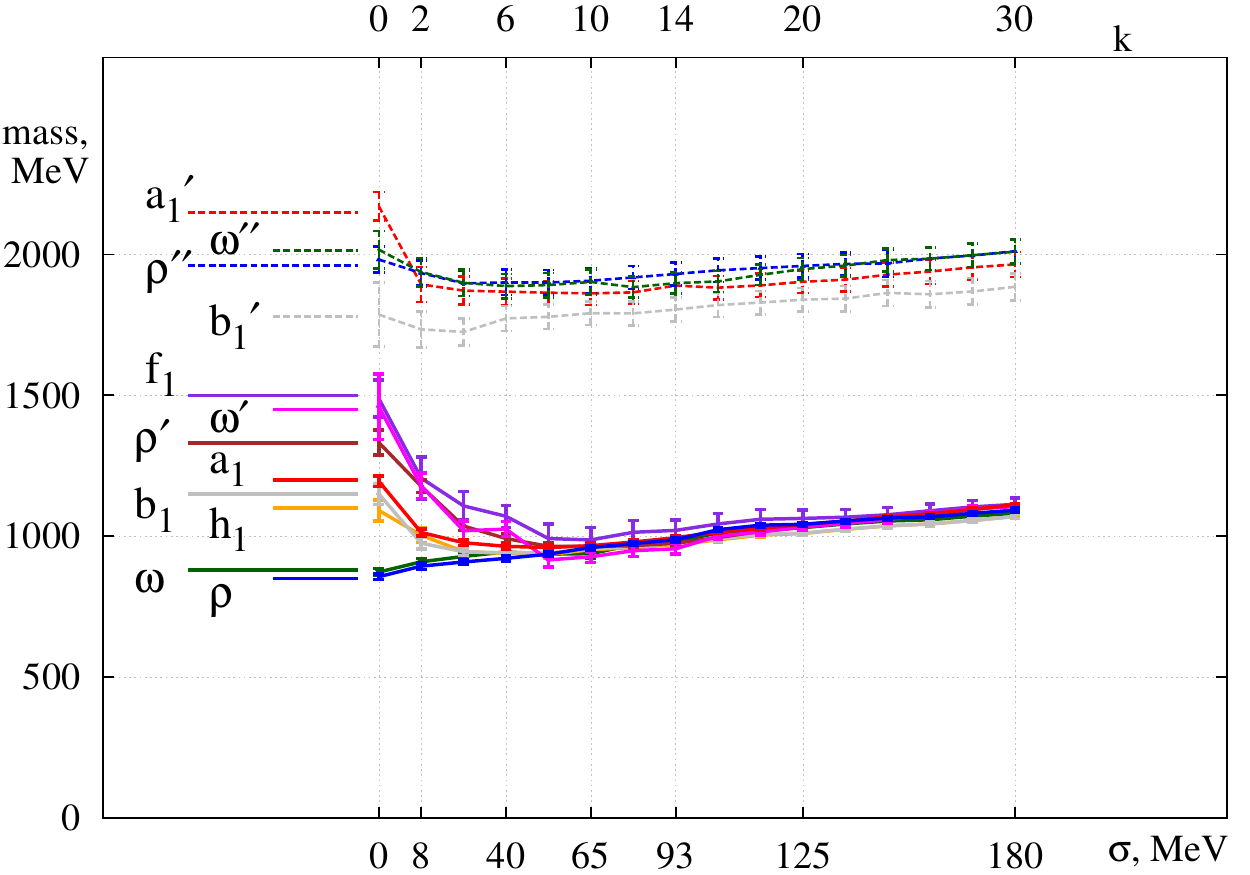}\hfill\\ 
   \caption{Mass evolution of $J=1$ mesons on exclusion of the quasi-zero modes. The value $\sigma$
denotes the energy gap.}\label{fig:allJ1}
\end{figure*}

All final results for masses, obtained from the one-exponential fits to the eigenvalues of
the cross-correlation matrices are shown in Fig. \ref{fig:allJ1}. The fit ranges, the extracted
masses with their errors are given in the Appendix B in Table \ref{tab:fitsk0}.   

We see a degeneracy in all possible chiral multiplets of the $J=1$ mesons (see Fig.
\ref{fig:symj1}): $\rho - a_1$, $b_1 -
\omega^\prime$ and $h_1 - \rho'$. This evidences a restoration of $SU(2)_L \times SU(2)_R$.
Simultaneously
the $U(1)_A$ symmetry gets also restored, because all four mesons $h_1, \rho', b_1, \omega^\prime$
are degenerate. A degeneracy of all eight $J=1$ mesons indicates a restoration of an even higher
symmetry $SU(4) \supset SU(2)_L \times SU(2)_R \times
U(1)_A$ \cite{G}. The  transformations of the latter mix the
components of the fundamental
four-component vector $(u_L,u_R,d_L,d_R)$. The high degeneracy of the
energy levels and the $SU(4)$ symmetry imply absence of a color-magnetic
field in the system and could be interpreted as a manifestation of a dynamical  string in \QCDbar.

\section{ Conclusion}

We have studied  all possible $\bar{q} q$, $J=0,1$ mesons in a dynamical lattice simulation with the
manifestly chirally invariant overlap Dirac operator and their behavior upon reduction of the lowest-lying eigenmodes
of the Dirac operator from the valence quark propagators. In the  $\pi,\sigma,a_0,\eta$ channels ($J=0$)    we
observe a simultaneous restoration of both chiral and $U(1)_A$ symmetries: All possible
point-to-point correlators become identical. The ground states of $\pi,\sigma,a_0,\eta$ mesons do
not survive the unbreaking, however.
They disappear from the physical spectrum of the bound states, because
the corresponding correlation functions decay not exponentially. 

In the $J=1$ channels
we find in contrast a very clean exponential decay of the correlators. Consequently, the
$J=1$ states survive the unbreaking of the chiral symmetry. After removal of the quasi-zero modes
their masses become manifestly chirally symmetric. 

In the  present case we also have evidence of
a simultaneous restoration of both $SU(2)_L \times SU(2)_R$ and $U(1)_A$ symmetries, like in the
$J=0$ channels. In the $J=1$ channels we clearly see a degeneracy of all eight possible mesons,
which signals a restoration of the higher symmetry, the $SU(4)$. The latter symmetry is not a
symmetry of the QCD Lagrangian, but is an emergent symmetry that appears from the QCD dynamics upon
removal of the close-to-zero Dirac modes. It operates only in $J \geq 1$ mesons. This symmetry
suggests an absence of  a color-magnetic field and can be
interpreted as a symmetry
of a dynamical string in \QCDbar.

\begin{appendix}

\section{Correlation functions}\label{sec:appA}

In this section we present an outline of the meson two-point correlation function computation with
stochastic estimation techniques. Here is a list of quark vectors representing quark
sources and solution vectors according to \cite{Aoki:2009qn,Foley:2005ac},
\begin{equation}
\begin{split}
 \{ v_k \}=&  \Big\{u_1,u_2,...,u_{N_{e}},x_1,...,x_{N_d}\Big\} \\
  \{  w_k \}=&  \Big\{\frac{u_1}{\lambda_1},\frac{u_2}{\lambda_2},...,
            \frac{u_{N_{e}}}{\lambda_{N_{e}}}
            ,P_l\eta_1,...,  P_l\eta_{N_d}\Big\},
\end{split} 
 \end{equation}
 where $u_{k}$ are $N_e=100$ low-lying modes and $\eta_{N_d}$ single $Z(2)$ noise vector for each
configuration. These are diluted into $N_d=3\times4\times N_t/2$ vectors $\eta_d$ ($d = 1,
..,Nd)$, which have nonzero elements only for a single combination
of color and Dirac indices and at two consecutive time slices. The vectors $x_d$ represent the solution of the linear
equation 
\begin{equation}
D_{ov}(m)x_d = P_l\eta_d
 \end{equation}
with a low-mode projector $P_l=1-\sum_{n=1}^{N_e}u_n u_n^\dagger$. 
Next we use these vectors for the  evaluation of two-point functions in the isovector meson
channel, 
 \begin{equation}
 \begin{split}
 & G^{I=1}_{\Gamma}(t;\text{\bf{p}}=0) \\
 &=\;\langle(\bar{q_1}\Gamma q_2)(t)(\bar{q_2} \Gamma^\dagger q_1)(0)\rangle \\
 &=\sum_{n=1}^{N_e+{N_d}} \sum_{m=1}^{N_e+{N_d}} \mathcal{O}^{(m,n)}(t) 
\mathcal{O}^{(n,m)}(0)\\
 &=\; C(t), 
 \end{split}                                         
 \end{equation}
  and in the isoscalar meson channel,
 \begin{equation}
 \begin{split}
 & G^{I=0}_{\Gamma}(t;\text{\bf{p}}=0) \\
 &=\;\langle(\bar{q}\Gamma q)(t)(\bar{q} \Gamma q)(0)\rangle \\
 &= \;\;\, \sum_{n=1}^{N_e+{N_d}} \sum_{m=1}^{N_e+{N_d}} \mathcal{O}^{(m,n)}(t) 
\mathcal{O}^{(n,m)}(0)\\ 
 &- \; 2\sum_{n=1}^{N_e+{N_d}} \mathcal{O}^{(n,n)}(t)
\sum_{m=1}^{N_e+{N_d}} \mathcal{O}^{(m,m)}(0)\\
 &= \; C(t)+D(t),
 \end{split}                                         
 \end{equation}
with 
\begin{equation}
  \mathcal{O}^{(n,m)}(t)=\sum_{\text{\bf{r}}}\phi(\text{\bf{r}})
   w_m(\text{\bf{x}}+\text{\bf{r}},t)\Gamma v_n(\text{\bf{x}},t)
 \end{equation}
 defined as a smeared meson interpolating field at time $t$ with $\Gamma$ a product of Dirac
matrices and smearing
functions $\phi$. We used an exponential type of smearing functions and $\Gamma$ structure
combinations $\{\phi_s,\Gamma \}$  listed in Tables \ref{tab:smr} and \ref{tab:int} for $J=0, 1$
mesons.

 \begin{table}[h]
\vspace{5mm}
\begin{center}
\begin{ruledtabular}
\begin{tabular}{|c|l|r|l|} 
s  & $\phi_s(\text{\bf{r}}) \propto$& s & $\phi_s(\text{\bf{r}}) \propto$   \\  \hline
1  &$\delta_{\text{\bf{r}},0}$  & 8  & $|\text{\bf{r}}|e^{-|\text{\bf{r}}|}$\\
2  & const  &    9  &$|\text{\bf{r}}|e^{-|\text{\bf{r}}|^2}$ \\ 
3  & $e^{-0.4 |\text{\bf{r}}|}$ &   10  & $|\text{\bf{r}}|^2e^{-0.2|\text{\bf{r}}|^2}$\\ 
4  & $e^{-|\text{\bf{r}}|}$&   11  & $|\text{\bf{r}}|^2e^{-0.4|\text{\bf{r}}|^2}$\\
5  & $e^{-0.4 |\text{\bf{r}}|^2}$&  12  & $|\text{\bf{r}}|^2e^{-0.7|\text{\bf{r}}|^2}$\\ 
6  & $e^{-0.7 |\text{\bf{r}}|^2}$&  13  & $|\text{\bf{r}}|^2e^{-|\text{\bf{r}}|^2}$\\
7  & $e^{-|\text{\bf{r}}|^2}$&  14-26 & $\phi_s=\phi_{1-13}$ \\  
 \end{tabular}
 \end{ruledtabular}
\end{center}
 \caption{Smearing functions with normalization $\sum_r|\phi_s(\text{\bf{r}})|=1$}\label{tab:smr}
\end{table}

\section{Fit results}\label{sec:appB}

Single exponential effective mass fits and corresponding $\chi^2$/d.o.f. are presented in Table
\ref{tab:fitsk0}. It is possible to optimize the set of
interpolators in order to improve the signal of the extracted states and consequently $\chi^2$/d.o.f. ratios
at each truncation level $k$ of near-zero modes. For the sake of consistency, we present numerical
results for masses of mesons obtained with the fixed set of interpolators in each quantum channel for
all values of $k$.

\begin{table*}[htb]
\begin{center}
\begin{ruledtabular}
\begin{tabular}{|c|c|c|c|c|c|c|c|c|c|c|c|}
 \multicolumn{6}{|c|}{$k = 0$} & \multicolumn{6}{c|}{$k = 12$}\\\hline
state& $n$ & am   &$\chi^2$/d.o.f.&  $t$ &  $  \{s\}$&
state& $n$ & am   & $\chi^2$/d.o.f.&  $t$ & $  \{s\}$\\
\hline
\multicolumn{1}{|c|}{\multirow{3}{1cm}{$\;\rho$}} & 1 & 0.514  $\pm$  0.006  & 1.37/6  & 4
- 11 &
\multicolumn{1}{l|}{\multirow{3}{1.7cm}{4,5,7,9,   13,17,18,  19,22,24}} &
\multicolumn{1}{c|}{\multirow{3}{1cm}{$\rho$}} & 1 & 0.585  $\pm$  0.005  & 13.49/8  & 4 - 13 &
\multicolumn{1}{l|}{\multirow{3}{1.7cm}{4,5,7,9, 13,17,18, 19,22,24}}\\
\multicolumn{1}{|c|}{ } & 2 & 0.801  $\pm$  0.027  & 11.12/4  & 3 - 8 &  &\multicolumn{1}{c|}{ } &
2 & 0.581  $\pm$  0.006  & 22.22/7  & 4 - 12 &\\
\multicolumn{1}{|c|}{ } & 3 & 1.191  $\pm$  0.028  & 1.66/4  & 3 - 8 &  &\multicolumn{1}{c|}{ } &
3 & 1.153  $\pm$  0.024  & 1.35/3  & 3 - 7 &  \\
\hline
\multicolumn{1}{|c|}{\multirow{3}{1cm}{$\;\omega$}} & 1 & 0.524  $\pm$  0.008  & 4.87/7  & 3 - 11 &
\multicolumn{1}{l|}{\multirow{3}{1cm}{4,5,7,9, 13,17,18, 19,22,24}}
&\multicolumn{1}{c|}{\multirow{3}{1cm}{$\omega$}} & 1 & 0.583  $\pm$  0.006  & 1.92/5  & 5 - 11 &
\multicolumn{1}{l|}{\multirow{3}{1cm}{4,5,7,9, 13,17,18, 19,22,24}}\\
\multicolumn{1}{|c|}{ } & 2 & 0.877  $\pm$  0.070  & 1.14/4  & 3 - 8 &  &\multicolumn{1}{c|}{ } &
2 & 0.570  $\pm$  0.012  & 0.38/4  & 5 - 10 &\\
\multicolumn{1}{|c|}{ } & 3 & 1.213  $\pm$  0.040  & 0.22/2  & 3 - 6 & &\multicolumn{1}{c|}{ } & 3
& 1.133  $\pm$  0.023  & 3.83/3  & 3 - 7 & \\
\hline
\multicolumn{1}{|c|}{\multirow{2}{1cm}{$\;a_1$}} & 1 & 0.718  $\pm$  0.012  & 0.12/5  & 3 - 9 &
\multicolumn{1}{l|}{\multirow{2}{1.7cm}{4,8,9,12,13}}& \multicolumn{1}{c|}{\multirow{2}{1cm}{$a_1$}}
&
1 & 0.589  $\pm$  0.005  & 21.91/8  & 4 - 13 & \multicolumn{1}{l|}{\multirow{2}{1cm}{4,8,9,12,13}}\\
\multicolumn{1}{|c|}{ } & 2 & 1.305  $\pm$  0.030  & 2.86/3  & 3 - 7 &  &\multicolumn{1}{c|}{ } &
2 & 1.121  $\pm$  0.025  & 2.11/4  & 4 - 9 &\\
\hline
\multicolumn{1}{|c|}{\multirow{1}{1cm}{$\;f_1$}} & 1 & 0.894  $\pm$  0.039  & 1.00/4  & 3 - 8 &
\multicolumn{1}{l|}{\multirow{1}{1cm}{5,7,9,11,12}} &\multicolumn{1}{c|}{\multirow{1}{1cm}{$f_1$}} &
1 & 0.609  $\pm$  0.025  & 0.13/2  & 5 - 8 & 5,7,9,11,12 \\
\hline
\multicolumn{1}{|c|}{\multirow{2}{1cm}{$\;b_1$}} & 1 & 0.691  $\pm$  0.023  & 0.67/3  & 3 - 7 &
\multicolumn{1}{l|}{\multirow{2}{1cm}{4,5,6,7,8,9}} &\multicolumn{1}{c|}{\multirow{2}{1cm}{$b_1$}} &
1 & 0.579  $\pm$  0.005  & 4.61/6  & 3 - 10 &
\multicolumn{1}{l|}{\multirow{2}{1cm}{4,5,6,7,8,9}}\\
\multicolumn{1}{|c|}{ } & 2 & 1.074  $\pm$  0.068  & 0.14/3  & 3 - 7 &  \multicolumn{1}{|c|}{ }& & 2
& 1.076  $\pm$  0.026  & 0.18/3  & 3 - 7 & \\
\hline
\multicolumn{1}{|c|}{\multirow{2}{1cm}{$\;h_1$}} & 1 & 0.656  $\pm$  0.022  & 0.22/3  & 3 - 7 &
\multicolumn{1}{l|}{\multirow{2}{1cm}{4,5,6,7,8,9}} &\multicolumn{1}{c|}{\multirow{2}{1cm}{$h_1$}} &
1 & 0.576  $\pm$  0.005  & 7.03/9  & 3 - 13 & \multicolumn{1}{l|}{\multirow{2}{1cm}{4,5,6,7,8,9}}\\
\multicolumn{1}{|c|}{ } & 2 & 1.064  $\pm$  0.074  & 0.17/2  & 3 - 6 &  &\multicolumn{1}{c|}{ } &
2 & 1.076  $\pm$  0.026  & 0.20/3  & 3 - 7 &\\
\end{tabular}
\end{ruledtabular}
\end{center}
\caption{Results of fits to the eigenvalues at a truncation levels $k=0,12$ for $J=1$ mesons: the
states are obtained from smeared interpolating fields. Each of the quantum states is denoted by
$n=1,2,...\,$. The mass values are in lattice units; $t$ denotes the fit range and $\{s\}$ labels
the set of smearing functions (see Table \ref{tab:smr} of Appendix A) used in the construction of
interpolating fields with respect to each quantum channel.}\label{tab:fitsk0}
\end{table*}

\end{appendix}
\acknowledgments
We are deeply grateful to S. Aoki, S. Hashimoto and T. Kaneko for their
suggestion to use the JLQCD overlap gauge configurations and quark propagators,
for their help and hospitality during our visit to KEK.
The calculations were performed on computing clusters of the University of Graz (NAWI Graz). Support from the
Austrian Science Fund (FWF) through the grants DK W1203-N16 and 
P26627-N16 is acknowledged.



\begin{thebibliography}{99}
\bibitem{myfootnote}
  We have checked that a nonlocality induced by such a procedure
  is, however, tiny \cite{SDGL}.
\end{thebibliography}
\end{document}